\documentclass[sn-aps,iicol]{sn-jnl}

\usepackage{breqn}

\jyear{2023}%

\theoremstyle{thmstyleone}%
\theoremstyle{thmstyletwo}%

\theoremstyle{thmstylethree}%

\begin{document}

\title[Article Title]{Viscous attenuation of gravitational waves propagating through an inhomogeneous background}


\author*[1]{\fnm{Shashank Shekhar} \sur{Pandey}}\email{shashankpandey7347@gmail.com}

\author[1]{\fnm{Arnab} \sur{Sarkar}}\email{arnab.sarkar14@gmail.com}

\author[2]{\fnm{Amna} \sur{Ali}}\email{amnaalig@gmail.com}

\author[1]{\fnm{Archan S.} \sur{Majumdar}}\email{archan@bose.res.in}

\affil[1]{\orgdiv{Dept. of Astrophysics $\&$ High Energy Physics}, \orgname{S. N. Bose National Centre for Basic Sciences}, \orgaddress{\street{Block - JD, Sector - III, Salt lake}, \city{Kolkata}, \postcode{700106}, \state{West Bengal}, \country{India}}}

\affil[2]{ \orgname{R-square RiskLab (RsRL)}, \orgaddress{\city{Dubai}, \country{UAE}}}



\abstract{We consider the propagation of gravitational waves in the late-time Universe in the presence of matter distribution inhomogeneities, and we also consider the cosmic fluid to be viscous. In this work, we investigate the cumulative effect of inhomogeneities and viscosity of the cosmic-fluid on the observables associated with the sources of the gravitational waves. Employing Buchert's averaging procedure in the backreaction framework, we consider a model of spacetime in which matter is distributed in-homogeneously across space. Using the modified redshift versus distance relation, through the averaging process in the context of the model, we study the variation of the redshift-dependent part of the observed gravitational wave amplitude for different combinations of our model parameters while simultaneously considering  damping of the gravitational wave amplitude due to viscosity of the cosmic-fluid. Then, we investigate the differences occurring in the variation of the redshift-dependent part of the observed gravitational wave amplitude due to consideration of viscous attenuation. We show that there are significant deviations after the inclusion of viscous attenuation in our analysis, depending on the chosen value of the coefficient of viscosity. Our result signifies the importance of the effect of viscosity, within the model of an inhomogeneous Universe, on precision measurements of parameters of compact-binary sources of gravitational waves.}





\maketitle

\section{Introduction }\label{sec:intro}

\par The century-old prediction of the possible existence of Gravitational waves (GWs) in Einstein's theory of General Relativity \cite{Einstein1, Einstein2} has recently found confirmation from the Laser Interferometer Gravitational-Wave Observatory (LIGO), and Virgo scientific collaborations \cite{Abbott_et_al, Ligo-virgo, Ligo-virgo2, Ligo-virgo3, Ligo-virgo4, Ligo-virgo5}, which has opened a new window to decipher the mysteries of Universe. With more and more GW data pouring in, one expects that the GWs will provide more insight into diverse phenomena, such as the origins of black holes, the extreme conditions inside neutron stars, the chronicle of how the Universe structured itself into galaxies, the physics of the first few moments in the aftermath of the Big-bang and the standard picture of Universe itself.

\par Since gravitational wave observations are used to infer various fundamental
properties related to the source of emission, it is important to have a complete understanding of the physics of their propagation from the source to us through the intervening background which in the present Universe is dominated by the dark components, {\it viz.}, dark matter and dark energy. Properties of the cosmic fluid could be cause for significant attenuation of the amplitude of GWs propagating through it. It may be noted that though electromagnetic (EM) waves are not affected by the viscosity of matter, GWs are indeed affected by viscosity \cite{PRASANNA1999120, Hawking,  Esposito, Madore1973, Anile1978, Goswami, brevik}. GWs have to work against the viscous matter while passing through it, resulting in loss of its energy, which is manifested in attenuation of amplitude or damping of the GWs.\footnote{It may be mentioned that similar damping occurs in EM waves when they traverse through any matter having sufficiently high conductivity.} Studies \cite{Nahuel} have been done analyzing the interaction between a viscous fluid (the primordial plasma in this case) and the primordial gravitational waves using a relativistic hydrodynamic theory. There also exist studies \cite{Moretti2020, universe7120496} describing damping of GWs due to non-collisional media during the propagation in alternate gravitational theories. For example, it is  possible for the longitudinal scalar modes of the GWs from Horndeski theories, to be damped by the non-collisional ensemble of massive particles.

\par Cosmologists have used viscosity in wide-ranging studies over the years. The initial singularity at the big bang can be avoided by invoking shear, and bulk viscosity \cite{PADMANABHAN1987433, Murphy}. Viscosity has  been used to explain dark energy \cite{Fabris2006, Gagnon_2011, Floerchinger}, and it has been shown that the Universe’s accelerated expansion can be due to the effect of viscosity \cite{ PADMANABHAN1987433, Gagnon_2011, Floerchinger, Atreya_2018, Mohan2017, Das2012, brevik2}. In certain  other schemes, the neutrino mass \cite{Anand_2018} and the 21-cm emission temperature \cite{Halder_2022} can also be constrained using viscous cosmology. The viscous matter in the path of propagation of GWs could be most likely in the form of certain types of dark matter  \cite{ Atreya_2018, Natwariya2020, Spergel, TULIN20181, Kaplinghat}, though even some of the visible stellar matter may also have some amount of viscosity.  Such dark matter with dissipative dynamics induced by viscosity can possibly settle the strain between Planck Cosmic Microwave Background (CMB) and Large Scale Structure (LSS) observations \cite{Anand_2017}. GWs in the presence of viscosity have  been suggested as probes of such viscous cosmological models \cite{Goswami}.  
 
\par Cosmological observations have also revealed that though the Universe is smooth and uniform on the very largest of scales, given the standard picture of the Big Bang and the known age of the Universe, this is not true for smaller scales. The transition from homogeneity to inhomogeneity at smaller scales has been indicated by various tests from cosmological observations like  the WiggleZ Dark Energy Survey \cite{WiggleZDE}. Analysis based on Sloan Digital Sky Survey \cite{Sylos_Labini_2009} have revealed absence of homogeneity in the large-scale galaxy distribution. Current estimates to analyze large-scale fluctuations in the luminous red galaxy samples based on higher-order correlations have found significant (more than 3 $\bar{\sigma}$) deviations from the $\Lambda$ cold dark matter ($\Lambda$CDM) mock catalogues on samples as large as 500 $h^{-1}$ Mpc \cite{wiegand_scale}. Thus, inhomogeneities due to structures may have important effects on length scales even as large as 500 $h^{-1}$ Mpc. 

It has been argued that local inhomogeneities may impact the overall evolution
of the Universe, through the backreaction arising from the process of 
averaging \cite{Shirokov1998, Ellis1984, Futamase,  Zalaletdinov1992, Zalaletdinov1993, Buchert, Weigand_et_al}. The backreaction effect quantifies the non-linear process of structure formation on the mean global evolution of the Universe. Several works have been done to explain the accelerated expansion of the late-time Universe through backreaction \cite{Schwarz, Rasanen_2004, wiltshire, Kolb_2006}, though there exists a debate in the literature  \cite{Ishibashi_2005} whether inhomogeneities could account for the accelerated expansion. Recently, the Hubble tension \cite{Riess_2021, Freedman_2021} arising from a discrepancy in the inferred value of the Hubble parameter from local measurements compared to that from early Universe physics, has attracted a lot of attention.  It may be noted that backreaction induced 
curvature may possibly explain the larger values of the Hubble parameter
obtained locally \cite{Heinesen_2020}. 

Generally, GW analysis is done by deeming that GW propagates through a homogeneous and isotropic spacetime, described by an FLRW-metric. However, GW sources which are the subject of the present observations \cite{Abbott_et_al, Ligo-virgo, Ligo-virgo2, Ligo-virgo3, Ligo-virgo4, Ligo-virgo5}, lie well within the scale of 500 $h^{-1}$ Mpc. The analysis of the effect of inhomogeneities on the propagation of GWs may be of significance for precision measurements in the emerging field of GW astronomy. 

The motivation of the present analysis is to study the propagation of GWs
through the background containing viscous dark matter in the presence of
inhomogeneities. It has been  shown earlier that the inclusion of the effect of local inhomogeneities leads to a non-trivial impact on the propagation of EM waves in averaged spacetime \cite{Coley, R_s_nen_2009,  Gasperini_2011, Fleury, Fleury2, Fleury3_2014, Bagheri_2014}. The cosmological-distance versus redshift relations gets modified due to averaging over inhomogeneities \cite{Futamase2,Koksbang_2019,Koksbang2},
leading to interesting prospects for the detection of signatures of inhomogeneities through observations of Hubble expansion \cite{Koksbang_PRL}. GWs act as fellow carriers of information to EM waves, and analysis pertaining to the former opens up a new avenue to the physics of the evolution of the Universe starting from early times, offering insight into the nature of gravity itself, and ranging to the current cosmic acceleration.

GWs have particular relevance for those sources which do not emit any EM signals. Here we consider compact objects in binary formations, e.g., black hole - black hole (BH-BH)
binaries, from which emitted GWs can be detected after traversing through the background viscous fluid. We consider the background dynamics  arising from the backreaction of inhomogeneities due to structure formation. Specifically, we employ the Buchert formalism \cite{Buchert, Weigand_et_al} to quantify the effect of backreaction. Buchert's approach on backreaction has been
analysed  earlier in various efforts to obtain concurrence with cosmological observations related to the current acceleration without resorting to dark energy \cite{Schwarz, Rasanen_2004, wiltshire, Kolb_2006, Coley, Koksbang_2019, R_s_nen_2008, bose,  Bose2013, Ali_2017}. It has been shown using the Buchert framework \cite{Pandey_2022} that the amplitude of GWs produced from  binaries could deviate substantially from that in the case of a homogeneous spacetime described by the $ \Lambda$CDM model.

In the present work, we  show that the local \textit{viscous-inhomogeneities} in the path of propagation of GWs may have a considerable impact on the GW observables. The inclusion of viscosity affects the GW observables in ways different from the case of its absence \cite{Pandey_2022}. In the context of a 
simplified two-partitioned model of inhomogeneities within the context of the Buchert framework, here we  evaluate the attenuation of GWs resulting from our model in comparison with the standard analysis of the $\Lambda$CDM model. Our approach clearly brings out the quantitative differences in the GW signal
due to the inclusion of effects of viscosity and inhomogeneities, in combined as well as separate ways.

The paper is organized as follows. A brief description of the background dynamics is provided in \autoref{sec:LCDM}. Here we first discuss  a viscous $\Lambda$CDM model and next describe our inhomogeneous two-partitioned model (for both viscous and non-viscous cases) within Buchert's averaging formalism.  In \autoref{sec:distvsz}, the modification of the redshift-distance relation due to the averaging procedure is presented. In \autoref{sec:GWamp}, the effect of local viscous inhomogeneities on the redshift-dependent part of GW amplitude is demonstrated. Finally, we present concluding remarks in \autoref{sec:conclusion}. 


\section{\label{sec:LCDM} Background dynamics}

For $\Lambda$CDM model (without viscosity), the Hubble parameter is given by the Friedmann equation
\begin{equation}
    H^2(a) = H^2_0\left(\frac{\Omega_{m0}}{a^3}+\frac{\Omega_{r0}}{a^4}+\frac{\Omega_{k}}{a^2}+\Omega_{\Lambda}\right) \label{eq:friedmann}
\end{equation}
where a is the scale factor, $H_0$ is the present value for the Hubble parameter, $\Omega_{m0}$ denotes the fractional matter density components (assumed pressure less) of the Universe, $\Omega_{r0}$ is the fractional radiation density term, $\Omega_{k}$ is the term related to the curvature, and $\Omega_{\Lambda}$ denotes the cosmological constant component.
 In practice, the contribution of the radiation at late times (i.e., at the time of structure formation) is negligible compared to the matter and cosmological constant terms. Also, observations indicate that the geometry of the Universe is almost flat, {\it viz.}, $\Omega_{k} \approx 0$.

\subsection{\label{subsec:vLCDM}  Viscous \texorpdfstring{$\Lambda$CDM model (v$\Lambda$CDM)}{ΛCDM model}}

GW  may propagate through dark matter in its path from its source to the observer. There are various theoretical models of dark matter. One of these is the Self Interacting Dark Matter (SIDM) model \cite{Natwariya2020, Spergel, TULIN20181, Kaplinghat}. In this model, self-interaction is introduced between the dark matter particles, which results in dissipation in the dark matter fluid. The outcome of this dissipation is the introduction of coefficients of shear, and bulk viscosities \cite{Atreya_2018}.
In our approach, the dark matter behaves as a viscous/dissipative component. The general structure of
this model is given by the field equation \cite{Weinberg}
\begin{equation}
    R_{\mu\nu}-\frac{1}{2}g_{\mu\nu}R+\Lambda g_{\mu\nu} = 8\pi GT_{\mu\nu}
\end{equation}\label{eq:Einsteineq}
where $R_{\mu\nu}$ represents the Ricci tensor, $R$ represents the Ricci scalar, $g_{\mu\nu}$ is the metric tensor, $\Lambda$ is the cosmological constant and $T_{\mu\nu}$ stands for the energy-momentum tensor of the viscous matter. This tensor possesses both the perfect fluid structure as well as the possible dissipative effects  such that \cite{Pimentel2016, Weinberg}
\begin{equation}
    T^{\mu\nu}=pg^{\mu\nu}+(p+\rho)u^{\mu}u^{\nu}+\Delta T^{\mu\nu}
\end{equation}\label{eq:Tmunu}
where $\rho$ is the density, $p$ is the pressure and the component $\Delta T^{\mu\nu}$ is the viscous contribution to the fluid,
\begin{equation}
    \Delta T^{\mu\nu} = -2\eta \sigma^{\mu\nu} - \xi\Theta (g^{\mu\nu} + u^\mu u^\nu)
\end{equation}\label{eq:deltaTmunu}
Here $\xi$ is the bulk viscosity, $\eta$ is the shear viscosity, $\Theta = u^\mu_{;\mu}$ is the expansion, $u^{\mu}$ is the 4 - velocity and ";" represents the covariant derivative.
For simplicity, we set the pressure, $p$ = 0. Then, our dark matter possesses only the viscous pressure given by \cite{Barbosa_2017}
\begin{equation}
    p_v = -\xi u^\mu_{;\mu} \label{eq:pv}
\end{equation}
 In the FLRW metric, the bulk viscous pressure reduces to
\begin{equation}
    p_v = -3H\xi \label{eq:pv2}
\end{equation}

 Dark matter physics can incorporate some possible dissipative mechanisms \cite{Atreya_2018, Natwariya2020, Spergel, TULIN20181, Kaplinghat,  Anand_2017}. Only the bulk viscosity remains compatible with the assumption of large-scale homogeneity and isotropy. The other processes, like shear and heat conduction, are directional mechanisms that decay as the Universe expands.
Shear viscosity has mostly been neglected in these studies on the grounds of not contributing to a homogeneous and isotropic universe, which is undoubtedly true at the large-scale background level \cite{Barbosa_2017, Velten_2014}. Hence, for our purpose, for this viscous $\Lambda CDM model$, shear viscosity does not contribute to the dynamics of an isotropic and homogeneous background. However, shear viscosity does indeed play a role
in the attenuation of gravitational waves, as we will see later in \autoref{sec:GWamp}.

Now, for a viscous $\Lambda$CDM model, using the FLRW metric, the Friedmann equation reads,
\begin{equation}
    H^2 \equiv \left(\frac{\dot{a}}{a}\right)^2 = \frac{8\pi G}{3}\rho_v+\frac{\Lambda}{3}. \label{eq:Friedmann2}
\end{equation}
Here, $\rho_v$ stands for the density of viscous matter and denotes all the matter components. We have assumed that all the matter components are endowed with viscous properties. For our purpose here, a proper separation between baryons and dark matter is unnecessary. One should also recall that baryons 
contribute about $1/6^{th}$ of the present total matter distribution; thus, this is not expected to lead to appreciable changes in our analysis, as baryonic matter is a subdominant component in comparison to dark matter.
 By defining the fractional densities $\Omega_{v} = 8\pi G\rho_v/(3H_0^2)$ and $\Omega_{\Lambda} = \Lambda/(3H_0^2)$, where $H_0$ is the present value for the Hubble parameter, the Friedmann equation (\autoref{eq:Friedmann2}) becomes
\begin{equation}
    H^2 = H_0^2(\Omega_{v}+\Omega_{\Lambda}) \label{eq:H}
\end{equation}
Using now the fluid equation for $\rho_v$, one gets
\begin{equation}
    \dot{\rho}_v+3H(\rho_v+p_v)=0.\label{eq:rho}
\end{equation}
Using (\autoref{eq:pv2}), (\autoref{eq:rho}) can be recast as an equation for the fractional density $\Omega_v$ as
\begin{equation}
    a\frac{d\Omega_v}{da}+3\Omega_v(1+\omega_v) = 0 \label{eq:omegav}
\end{equation}
where we have defined the fluid equation of state parameter for the viscous dark matter fluid, $\omega_v$, as
\begin{equation}
    \omega_v \equiv \frac{p_v}{\rho_v}= -\frac{3H\xi}{\rho_v} \label{eq:EOS}
\end{equation}
Using this formalism, $\Omega_v$ as a function of the red-shift $z$ is calculated. The corresponding quantity in the $\Lambda$CDM case is $\Omega_{m0}(1+z)^3$ (\autoref{eq:friedmann}).


 \subsection{Buchert's formalism in a two-partitioned model}\label{subsec:Buchert_toy}
 
 A  popular approach for studying the effect of inhomogeneities is based on an averaging framework, and several averaging techniques have been proposed \cite{Shirokov1998, Ellis1984, Futamase,  Zalaletdinov1992, Zalaletdinov1993, Futamase2}. Since the Einstein equations are non-linear,  the solutions for an overall homogeneous matter distribution differ from the averaged solution for a general locally inhomogeneous matter distribution. In other
 words, the evolution of the homogeneous Universe at large scales may be slightly different from that of an inhomogeneous Universe, even though inhomogeneities, when averaged over a sufficiently large scale, might be negligible. The difference between the evolution of these models of the Universe gives the backreaction effect. It quantifies the non-linear effect of structure formation on the mean global evolution of the Universe. 
 
 In the averaging framework, \cite{Buchert}, the problem is simplified and restricted to scalar quantities only. Einstein equations are decomposed into a set of dynamical equations for scalar quantities. Averages on flow-orthogonal spatial hypersurfaces (vorticity is assumed zero) are defined as proper volume averages. Under certain assumptions, this leads to Buchert’s equations with a kinematical backreaction term \cite{Buchert}. Such an approach has  sparked considerable interest 
 as it has been shown that backreaction could lead to an agreement with cosmological observations without resorting to dark energy \cite{Schwarz, Rasanen_2004, wiltshire, Kolb_2006, Coley, Koksbang_2019, R_s_nen_2008, bose, Bose2013,  Ali_2017}. For details of Buchert’s averaging procedure, one may refer to Ref.~\cite{Buchert, Weigand_et_al}.
 Here we provide a brief overview of Buchert's formalism  required in
 context of the present analysis.  In Buchert's averaging scheme for scalars, averages of scalar quantities on flow-orthogonal spatial hypersurfaces are defined as 
\begin{equation} \label{eq:favg}
\langle f(t, x^i)\rangle_D := \frac{\int_D d^{3}x \sqrt{det(g_{ij})} f(t, x^i) }{\int_D d^{3}x \sqrt{det(g_{ij})} } \, , 
\end{equation}
where D is a spatial domain. One fundamental quantity characterizing this domain is its volume, which is given by,
\begin{equation}\label{eq:VD}
    V_D(t) := \int_D d^3x  \sqrt{det(g_{ij})} \, .
\end{equation}
The normalized dimensionless effective volume scale factor $a_D$ is defined by
\begin{equation}\label{eq:aD}
    a_D (t) := \left(\frac{V_D(t)}{V_{D_0}}\right)^{1/3} \, , 
\end{equation}
which is normalized by the volume $V_{D_0}$ of the domain $D$ at some reference time $t_0$ which we can take as the present time.

Spatially averaging the Raychaudhuri equation, the Hamiltonian constraint and the continuity equation, one obtains the  equations for effective scale factor in Buchert's formalism, which are respectively,
\begin{equation}\label{eq:Rayeqn}
    3\frac{\ddot{a}_D}{a_D} = -4\pi G\langle\rho\rangle_D + Q_D + \Lambda \, ,
\end{equation}
\begin{equation}\label{eq:HamiltonianCeqn}
    3H_D^2 = 8\pi G\langle\rho\rangle_D - \frac{1}{2}\langle R\rangle_D - \frac{1}{2}Q_D + \Lambda \, ,
\end{equation}
\begin{equation}\label{eq:continuity}
    \partial_t\langle\rho\rangle_D + 3H_D\langle\rho\rangle_D = 0 \, , 
\end{equation}
where local averaged matter density $\langle\rho\rangle_D $,
averaged spatial Ricci scalar $\langle R\rangle_D$ and the Hubble parameter $H_D$ are domain dependent and are functions of time. $\Lambda$ is the cosmological constant ($\Lambda = 0$ for our model and for our purpose here). $Q_D$ is called the backreaction term which quantifies the averaged effect of the inhomogeneities in the domain $D$ and is defined as
\begin{equation}\label{eq:QD}
    Q_D := \frac{2}{3}\left(\langle \theta^2\rangle_D-\langle\theta\rangle_D^2\right) - 2\langle\sigma^2\rangle_D \, ,
\end{equation}
where $\theta$ is the local expansion rate and $\sigma^2 := \frac{1}{2}\sigma^i_j\sigma^j_i$ is the shear-scalar. $Q_D$ is zero for a homogeneous domain. The departure from homogeneity is ingrained in this term. $Q_D$ and $\langle R\rangle_D$ are inter-related by the equation:
\begin{equation}\label{eq:condition}
    \frac{1}{a_D^2}\partial_t(a_D^2\langle R\rangle_D) + \frac{1}{a_D^6}\partial_t(a_D^6Q_D) = 0 \, .
\end{equation}
(\autoref{eq:condition}) couples the time evolution of the backreaction term with the time evolution of averaged intrinsic curvature and signifies the departure from FLRW-cosmology, where there is no such coupling. 
\par
In this framework, the domain $D$ is partitioned into non-interacting subregions $\mathcal{F}_l$ composed of elementary space entities $\mathcal{F}_l^{(\alpha)}$. Mathematically, $D = \cup_l\mathcal{F}_l$, where $\mathcal{F}_l = \cup_{\alpha}\mathcal{F}_l^{(\alpha)}$ and $\mathcal{F}_l^{(\alpha)}\cap\mathcal{F}_m^{(\beta)} = \emptyset$ for all $\alpha\neq\beta$ and $l\neq m$. The average of any scalar function $f$ on the domain $D$ is given by,
\begin{equation}\label{eq:fD_sum}
\begin{split}
    \langle f \rangle_D &= V_D^{-1}\int_D f  \sqrt{det(g_{ij})} d^{3}x \\
    &= \sum_l V_D^{-1}\sum_{\alpha}\int_{\mathcal{F}_l^{(\alpha)}} f  \sqrt{det(g_{ij})} d^{3}x \\
    &= \sum_l \frac{V_{\mathcal{F}_l}}{V_D}\langle f\rangle_{\mathcal{F}_l} = \sum_l \lambda_l\langle f\rangle_{\mathcal{F}_l} \, , 
\end{split}   
\end{equation}
where $\lambda_l = V_{\mathcal{F}_l}/V_D$ is the volume fraction of the subregion $\mathcal{F}_l$ such that $\sum_l \lambda_l = 1$ and $\langle f\rangle_{\mathcal{F}_l}$ is the average of $f$ on the subregion $\mathcal{F}_l$. The above equation governs the averages of scalar quantities $\rho$, $ R$ and $H_D$. But $Q_D$, due to the presence of $\langle\theta\rangle_D^2$ term, does not follow the above equation. Instead, the equation for $Q_D$ is
\begin{equation}\label{eq:QD_sum}
    Q_D = \sum_l \lambda_lQ_l + 3\sum_{l\neq m}\lambda_l\lambda_m(H_l-H_m)^2 \, , 
\end{equation}
where $Q_l$ and $H_l$ are defined in the subregion $\mathcal{F}_l$ in the same way as $Q_D$ and $H_D$ are defined in the domain $D$ \cite{Weigand_et_al}.

We can also define scale factor $a_l$ for the individual subregions in the same way as $a_D$ has been prescribed for the domain $D$. Since, the domain $D$ comprises the different subregions $\mathcal{F}_l$ and all these subregions are disjoint, therefore $V_D = \sum_l V_{\mathcal{F}_l}$, which results in $ a_D^3 = \sum_l a_l^3$. Twice  differentiating this relation with respect to foliation time gives us,
\begin{equation}\label{eq:aD_sum}
    \frac{\ddot{a}_D}{a_D} = \sum_l\lambda_l\frac{\ddot{a}_l(t)}{a_l(t)}+\sum_{l\neq m}\lambda_l\lambda_m(H_l -H_m)^2 \, .
\end{equation}


Within the context of the above framework, we
 consider a two-partitioned model. In our model, the domain $D$, through which GWs propagate from the source to the observer, is an ensemble of two types of disjoint FLRW regions \cite{R_s_nen_2006}. These are - (i) the overdense region described by an FLRW region which can be assumed to be spatially flat, and (ii) the underdense region, which is a nearly empty FLRW region which is assumed to have a small density (compared to the overdense region) and negative spatial curvature. We consider the overdense  region to have viscous nature in our present work to study the effect of viscosity on the GW amplitude. 
Therefore for our two-partitioned model, (\autoref{eq:QD_sum}) effectively becomes,
\begin{equation}\label{eq:QDsum2}
    Q_{D} = \lambda_o Q_o + (1 - \lambda_o) Q_u + 6\lambda_o(1 - \lambda_o)(H_o - H_u)^2   \, , 
\end{equation}
where $\lambda_o$ denotes the volume fraction of the overdense region. Now, (\autoref{eq:condition}) couples the time evolution of the backreaction term $Q_D$ with the time evolution of the averaged 3-Ricci scalar curvature. Therefore, we can choose the curvatures of our individual sub-regions in such a way that the $Q_l$ term for these sub-regions becomes effectively zero \cite{Weigand_et_al}, i.e., $\mathcal{Q}_o = 0$ and $\mathcal{Q}_u = 0$. This has been done by taking the curvature of our overdense region to be zero, i.e., our overdense region is flat. On the other hand, we have assumed our underdense region to have Friedmann-like $a_u^{-2}$ constant curvature term. These assumptions along with (\autoref{eq:condition}) results in, $\mathcal{Q}_o = 0$ and $\mathcal{Q}_u = 0$. This stipulation to FLRW is an approximate assumption governing our model (in the more general case, the sub-domains may not necessarily be FLRW regions). From (\autoref{eq:QDsum2}), it can be seen that control over global backreaction can be achieved only if the individual backreaction terms are not set to zero.

In this work, we investigate the simultaneous impact of inhomogeneities and viscosity of the matter contained in the overdense region, on the amplitude of GWs for the same type of sources. For this, we consider two cases - (a) when the GW is assumed to propagate through a homogeneous and isotropic spacetime (with and without viscosity of associated matter), described by the FLRW metric, in the $\Lambda$CDM model, and (b) when the GW propagates through an inhomogeneous spacetime (both viscous and non-viscous cases), described by our model specified above.  
The physical interpretation of the underdense region in our model is that they represent the cosmic voids in the path of propagation of the GWs, and the overdense region represents all the matter content in that path. In this work, we have considered both scenarios (viscous and non-viscous) for the matter present in the overdense region of our model. A point to note here is that we are not considering viscous matter in addition to non-viscous matter here. The total matter content in both the non-viscous case and the viscous case is the same. It's just that in viscous case the matter content has viscous properties too. The overdense viscous region in our model portrays a viscous dark matter fluid.

\subsubsection{Non-viscous case (no bulk and shear viscosity)}\label{subsubsec:nonviscous}
Expressions for the scale factors for the two types of regions, overdense (non-viscous in this case) and underdense, for this model, are given as \cite{bose, Bose2013},
\begin{eqnarray}
    a_o = c_o t^{\alpha} ,\label{eq:aomod2}\\
    a_u = c_u t^{\beta}. \label{eq:aumod2}
\end{eqnarray}
Here $o$ represents the overdense region, and $u$ represents the underdense region. In this model, the scale factors of the two regions are proportional to cosmic time raised to some powers $\alpha$ and $\beta$ for over and under-dense regions, respectively. $\alpha$ varies from 1/2 to 2/3 since the evolution of $a_u$ is expected to be faster than that for the radiation-dominated case (1/2) and limited by the maximum value for the matter-dominated case (2/3). $\beta$ varies from 2/3 to 1 to denote any behaviour ranging from a  matter-dominated region ($ \beta = 2/3$) up to dark energy-dominated region ($ \beta  > 1 $). $c_u$ and $c_o$ are constants of proportionality and are given respectively as:
\begin{equation*}
\begin{aligned}
    c_o = \frac{a_{D_0}}{t_0^{\alpha}}, \\
    c_u = \frac{a_{D_0}}{t_0^{\beta}},
\end{aligned}    
\end{equation*}
where $a_{D_0}$ is the scale factor at the present time for the domain D. For present time $ t_{0} \approx 13.8 \, Gy $, $a_{D_0} = 1$ and $H_0 = 100$ $h$ km $sec^{-1}$ $Mpc^{-1}$. Therefore,
\begin{eqnarray}
    c_o = \frac{1}{t_0^{\alpha}},\label{eq:co_modparam} \\
    c_u = \frac{1}{t_0^{\beta}},\label{eq:cu_modparam}
\end{eqnarray}
Using (\autoref{eq:aD_sum}) for our model we get,
\begin{equation}\label{eq:aDsum_model}
    \frac{\ddot{a}_D}{a_D} = \lambda_o\frac{\ddot{a}_o}{a_o}+\lambda_u\frac{\ddot{a}_u}{a_u}+2\lambda_o\lambda_u(H_o -H_u)^2.
\end{equation}
Here $\lambda_o = V_{o}/V_D$ is the volume fraction for the overdense region, $V_o$ is the volume of the overdense region and  $\lambda_u$ is the volume fraction of underdense region such that $\lambda_u + \lambda_o = 1$. Now, using (\autoref{eq:aD}), $V_o$ can be written in terms of scale factor and initial volume fraction, $\lambda_o = \frac{a_o^3 V_{o_0}}{a_D^3 V_{D_0}}$, where $V_{o_0}$ is the volume of the overdense region at some reference time $t_0$. This in turn gives us,
\begin{equation}\label{eq:lambdao}
    \lambda_o = k_1\frac{t^{3\alpha}}{a_D^3}.
\end{equation}
where $k_1 = \frac{\lambda_{o_0}a^3_{D_0}}{t_0^{3\alpha}}$ is a constant and $\lambda_{o_0}$ , $a_{D_0}$ , $t_0$ are the present volume fraction of overdense region, the present global scale factor and the present time, respectively. (\autoref{eq:lambdao}) shows that the volume fraction of the overdense region is a function of $\alpha$ and $\beta$ (through $a_D$ as $a_D$ is a function of both $\alpha$ and $\beta$). Similarly, it can be shown that the volume fraction of the underdense region is a function of $\alpha$ and $\beta$. This implies that $(\alpha,\beta)$ governs the volume fractions of the 2 subregions in our 2-domain model. The present values of volume fractions of the 2 region are taken as $(\lambda_{o_0},\lambda_{u_0}) = (0.09,0.91)$ \cite{Weigand_et_al}.
Solving (\autoref{eq:aDsum_model}) gives us an expression for $a_D$, and using that expression in (\autoref{eq:Rayeqn}) gives us an expression of $\langle\rho\rangle_D$ for our model.

\subsubsection{Viscous case}\label{subsubsec:viscous}

 The sub-regions in our backreaction model are essentially FLRW regions. Hence, their dynamics are also governed by the standard Friedmann equations (\autoref{eq:friedmann}, \ref{eq:Friedmann2}). Only our overdense region has viscous matter contained in it. So, for our overdense region, the Friedmann equation reads (we have taken $\Lambda = 0$),
\begin{equation}
    {H_v}_o^2 \equiv \left(\frac{\dot{{a_v}_o}}{{a_v}_o}\right)^2 = \frac{8\pi G}{3}{\rho_v}_o. \label{eq:Friedmann2_viscous_overdense}
\end{equation}
Here, ${a_v}_o$ is the scale factor for our viscous overdense region, ${H_v}_o$ is the Hubble parameter for our viscous overdense region and ${\rho_v}_o$ stands for density of viscous matter in our overdense region and denotes all the matter components. We have assumed that all the matter components of our overdense region are endowed with viscous properties similar to the $\Lambda$CDM model with viscosity. Also, since our overdense region is an FLRW region (homogeneously and isotropically overdense), therefore here too the 
shear viscosity with coefficient $\eta$, doesn't have any effect on the dynamics of the background. The fluid equation for ${\rho_v}_o$ is given by,
\begin{equation}
    \dot{{\rho_v}_o}+3{H_v}_o({\rho_v}_o+{p_v}_o)=0.\label{eq:rhov_overdense}
\end{equation}
where ${p_v}_o$ is the bulk viscous pressure given by
\begin{equation}
    {p_v}_o = -3{H_v}_o\xi \label{eq:pv_overdense}
\end{equation}
By defining the fractional densities ${\Omega_v}_o = 8\pi G{\rho_v}_o/(3H_0^2)$, equation for the fractional density ${\Omega_v}_o$ is given as,
\begin{equation}
    a_o\frac{d{\Omega_v}_o}{da_o}+3{\Omega_v}_o(1+{\omega_v}_o) = 0 \label{eq:omegav_overdense}
\end{equation}
where we have defined the fluid equation of state parameter for the viscous dark matter fluid for our overdense region, ${\omega_v}_o$, as
\begin{equation}
    {\omega_v}_o \equiv \frac{{p_v}_o}{{\rho_v}_o}= -\frac{3{H_v}_o\xi}{{\rho_v}_o} \label{eq:EOS_overdense}
\end{equation}
Since (\autoref{eq:omegav_overdense}) is a first-order differential equation,  we require one boundary condition to solve it. We  take this boundary condition from our non-viscous backreaction model, given by
\begin{equation}
    {\Omega_v}_o(z=0)={\Omega}_o(z=0)\label{eq:boundary_condition}
\end{equation}
where ${\Omega}_o$ is the fractional density of the overdense region from the non viscous case.

The scale factor for our overdense region in our viscous backreaction model is calculated using (\autoref{eq:EOS_overdense}, \ref{eq:omegav_overdense} and \ref{eq:Friedmann2_viscous_overdense}). The boundary condition used for solving (\autoref{eq:omegav_overdense}) is (\autoref{eq:boundary_condition}). The ${a_v}_o$ that we get from the above-mentioned analysis is a function of $(\alpha,\beta)$. The scale factor for the underdense region, $a_u$ is still given by (\autoref{eq:aumod2}). The scale factor for domain D in this case ${a_v}_D$ is given by,
\begin{equation}\label{eq:aDmod2viscous}
    \frac{\ddot{a_v}_D}{{a_v}_D} = \lambda_o\frac{\ddot{a_v}_o}{{a_v}_o}+\lambda_u\frac{\ddot{a}_u}{a_u}+2\lambda_o\lambda_u({H_v}_o -H_u)^2
\end{equation}
The equation for the backreaction term in this case ${Q_v}_D$ is given by ($Q_u = 0; Q_o = 0$),
\begin{equation}
    {Q_v}_D =  6\lambda_o(1 - \lambda_o)({H_v}_o - H_u)^2, \label{eq:QD_Qo_Qu_viscous} 
\end{equation}
Solving (\autoref{eq:aDmod2viscous}) gives us an expression for ${a_v}_D$. (\autoref{eq:QD_Qo_Qu_viscous}) leads to an expression for ${Q_v}_D$. Using these expressions in (\autoref{eq:Rayeqn}) we get an expression of ${\langle\rho\rangle_v}_D$ for our  viscous backreaction model.  
The present value of the bulk viscosity parameter has been estimated in the
literature based on various theoretical considerations and observations
\cite{brevik, brevik2, brevik3, brevik_entropy}. By solving the energy conservation equation for bulk viscous flat Friedmann universes, the
 coefficient of bulk viscosity at present time $\xi$ has been
estimated to be $\sim 10^6$ Pa sec in \cite{brevik_entropy}. Other
analyses based on comparing the theoretical curves for $H = H(z)$ with  observations \cite{wang_meng}, and by studying the asymptotic behaviour in the equivalent phase space in a Friedmann model of the Universe with bulk viscous matter \cite{CUSAT_Sasidharan2016} have also been performed. In \cite{Velten_PRD}, a value of $\sim 10^7$ Pa sec for $\xi$ has been
suggested. Considering all the above studies, a reasonable range for
$\xi$ may be taken as $10^5$ Pa sec $<$ $\xi_0$ $<$ $10^7$ Pa sec.
For our present work, we have used the mid-range value of $10^6$ Pa sec  in the red-shift range $0\leq z\leq 5$. 

\begin{figure*}[h!]
    \includegraphics[scale = 0.75]{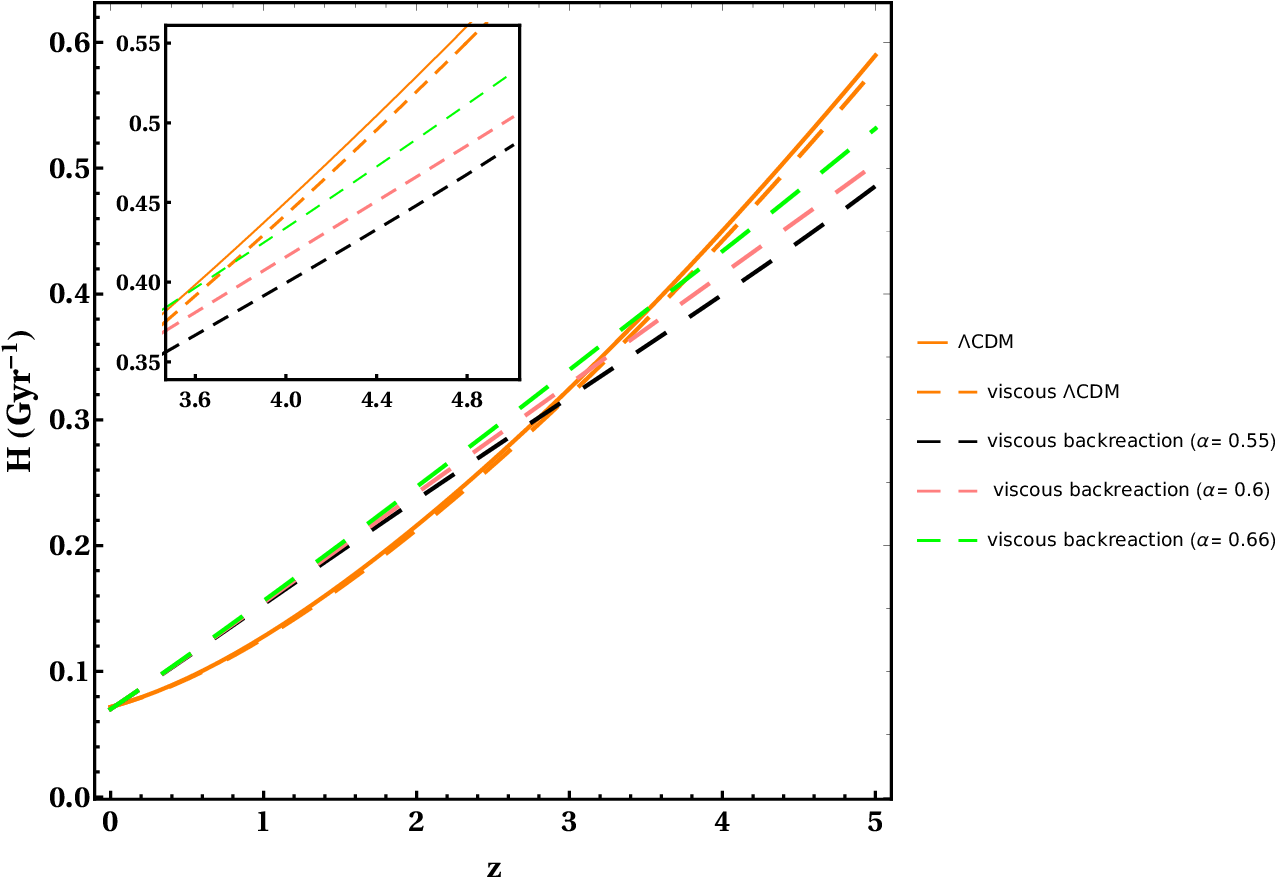}
    \caption{Plot of Hubble parameter vs redshift z for the $\Lambda$CDM model, viscous $\Lambda$CDM model and our viscous backreaction model. Plots for our viscous backreaction model are for varying values of $\alpha$ with fixed value of $\beta$ = 0.8. The value of $\xi$ used is $10^6$ Pa sec. The value of $H_0$  used is 100 $h$ km $s^{-1}$ $Mpc^{-1}$ $\simeq$ 0.07 $Gyr^{-1}$, with h = 0.7. Parameters used for $\Lambda$CDM are $\Omega_m$ = 0.31, $\Omega_\Lambda$ = 0.69. The inset shows the magnified portion of the plots at higher redshift.}
    \label{fig:Hv1}
\end{figure*}

In \autoref{fig:Hv1} we provide a plot of the Hubble parameter versus 
the redshift for $\Lambda$CDM, viscous $\Lambda$CDM and for our viscous backreaction model for various values of the backreaction model parameters. The underdense region in our model represents the cosmic voids in the path of propagation of the GWs, and the overdense region represents all the matter content in that path where the effect of viscosity is directly
evident. In \autoref{fig:Hv1}, for the plots of our viscous backreaction model, we fix the value of $\beta$ at 0.8 (representing a mid-range value in
the range of values of $\beta$ - (2/3, 1)), and we  vary the values of $\alpha$. 
As expected, for a fixed $\beta$, curves with larger value of $\alpha$ lead
to larger values of Hubble parameter. It can be seen that backreaction from inhomogeneities may lead to departure in the background Hubble evolution that may get accentuated for higher redshifts.


\vspace{0.29cm}   
\section{Redshift and distance relation }\label{sec:distvsz}

Buchert's averaging scheme provides us with a method of spatially averaging scalar quantities in the backreaction framework. Such quantities need to
be related to cosmological observables. A possible approach relies on the study of distance-redshift relation in an inhomogeneous universe  using an approximate metric \cite{Futamase2}. In our present investigation, we consider a more 
consistent scheme based on the procedure of averaging. The covariant scheme proposed by S. R\"{a}s\"{a}nen \cite{R_s_nen_2009,R_s_nen_4_2010} provides us with a way of doing this. This scheme gives the relation between effective redshift and angular diameter distance $D_A$ in the following way,
\begin{eqnarray} 
1+z = \frac{1}{a_{\mathcal{D}}} \, , \label{eq:covariantsch1} \\
H_{\mathcal{D}}\frac{d}{dz} \Big( (1+z)^{2} H_{\mathcal{D}} \frac{d D_{A}}{dz} \Big) = - \frac{4 \pi G}{c^{4}} \langle \rho_{\mathcal{D}} \rangle D_{A}   \, .  \label{eq:covariantsch2} 
\end{eqnarray}  
(\autoref{eq:covariantsch1}) provides an expression of effective redshift $z$ in terms of the scale factor $a_D$ of the domain $D$. Certain conditions must be satisfied to apply the covariant scheme. These are: (i) spatial averages are determined on hypersurfaces of statistical homogeneity and isotropy, and (ii) structure evolution is slow compared to the travel time of GW from source to the observer. 

For our model, domain D is the region of spacetime through which the GW travels while propagating from the source to the observer. Domain D could have any combination of fractions of underdense and overdense regions, i.e. any combination of $(\lambda_u, \lambda_o)$ as long as $\lambda_u + \lambda_o = 1$ is satisfied. $(\lambda_u, \lambda_o)$ in turn are governed by $(\alpha,\beta)$. In this work, we take various combinations of allowed values of $(\alpha,\beta)$ in our analysis. Using the expressions for $H_D$ and $\langle\rho_D\rangle$ that we calculated from our model for the two cases - viscous (with only $\xi$, but no $\eta$ since $\eta$ doesn't affect the background) and non-viscous, and using the covariant scheme (\autoref{eq:covariantsch1}, \autoref{eq:covariantsch2}), we can calculate $D_A$ for our model.

\begin{figure*}[h!]
    \includegraphics[scale = 0.74]{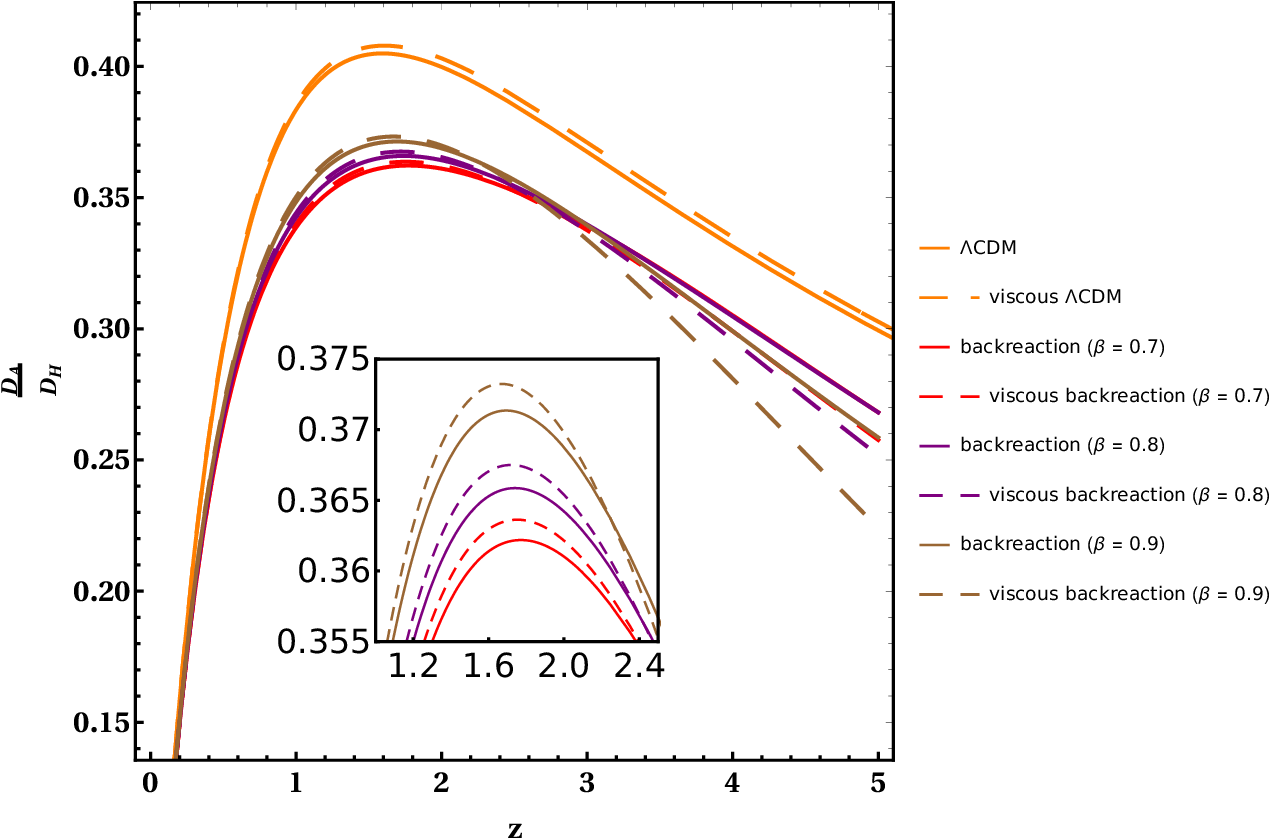}
    \caption{Plot of the ratio of angular diameter distance $D_{A}$ to the present Hubble length $D_{H}$ w.r.t. redshift, for the $\Lambda $CDM case and for our backreaction model with different combinations of $\beta$ and fixed value of $\alpha$ = 0.5 for both viscous and non-viscous cases. Value of $\xi$ used is $10^6$ Pa sec. Parameters used for $\Lambda$CDM are $\Omega_m$ = 0.31, $\Omega_\Lambda$ = 0.69. The inset shows a magnified portion of the plots for our backreaction model.}\label{fig:DA}
\end{figure*}

In (\autoref{fig:DA}), we plot the ratio of angular diameter distance for our model to the present Hubble-length ($D_{H}= cH_{0}^{-1} $, value of $H_0$ (present value of H) used is 100 $h$ km $s^{-1}$ $Mpc^{-1}$ $\simeq$ 0.07 $Gyr^{-1}$, with h = 0.7) as a function of effective redshift with different combinations of $(\alpha,\beta)$. We have studied earlier 
the effect of varying $\alpha$ on the dynamics (\autoref{fig:Hv1}). Here in \autoref{fig:DA} we explore the effect of varying $\beta$ on the angular diameter distance choosing a fixed value of $\alpha= 0.5$.  The value of $\xi$ used is $10^6$ Pa sec  in the range $0\leq z\leq 5$ \cite{brevik}. From this figure, one can see  that for low redshifts (z $<$ 0.5), curves for our model overlap with each other and with the $\Lambda$CDM curve, but as we increase the redshift, curves start deviating from the $\Lambda$CDM curve. Inclusion of viscosity in the analysis results in deviation in the plots for both the $\Lambda$CDM model and our model. For the $\Lambda$CDM case, plot of viscous case (dashed) has higher magnitude than the non-viscous case (solid). It is observed that for lower values of z $(z \lessapprox 2)$, plots of viscous cases (dashed curves) for our backreaction model have a higher magnitude of $D_A$ for the same value of $z$ in comparison to corresponding non-viscous cases (solid curves)  and for $z \gtrapprox 2$, the magnitude of $D_A$ for viscous cases (dashed curves) are smaller than those for the non-viscous cases (solid curves) for our model.

\begin{figure*}[h!]
    \centering
    \includegraphics[scale = 0.74]{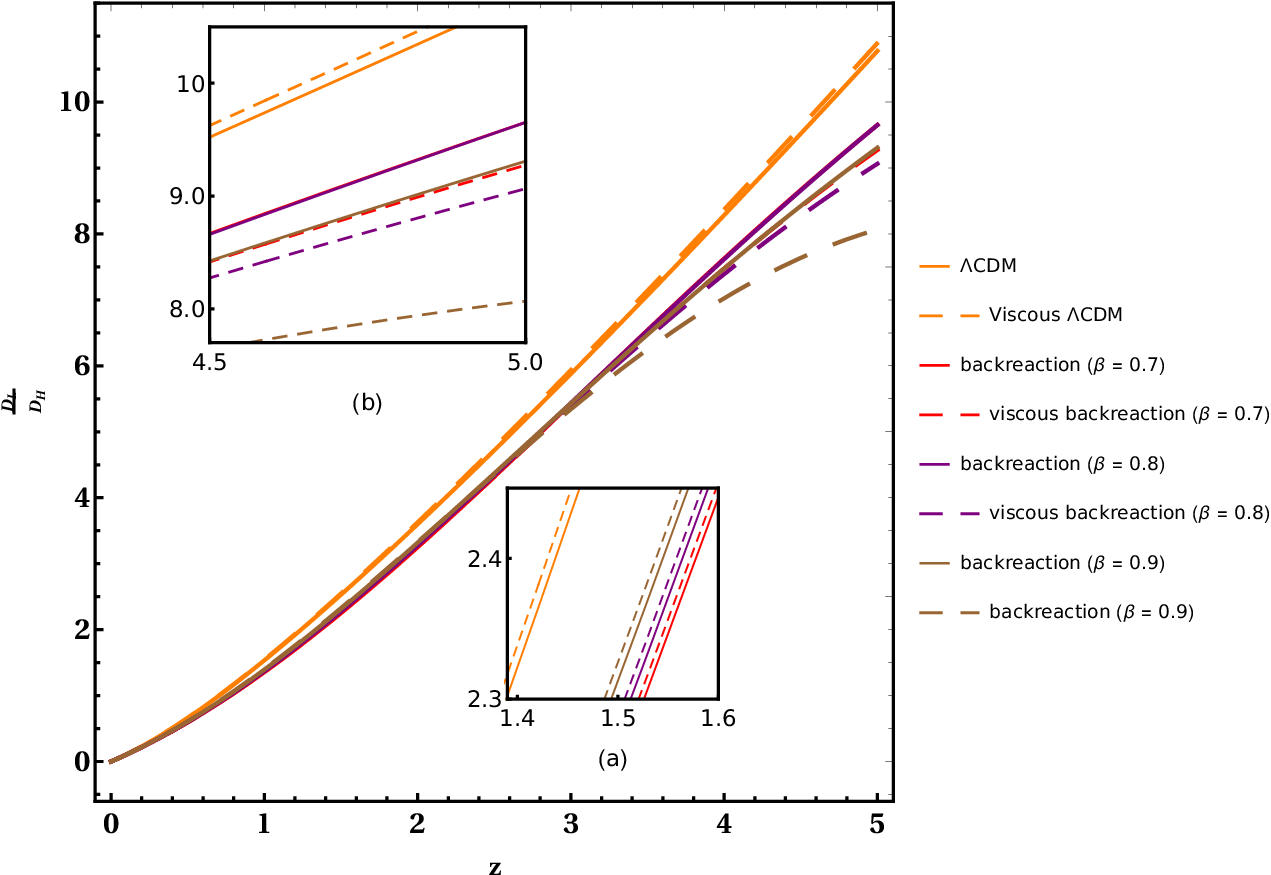}
    \caption{Plot of the ratio of luminosity distance $D_{L}$ to the present Hubble length $D_{H}$ w.r.t. redshift, for the $\Lambda $CDM case and for our backreaction model with different combinations of $\beta$ and fixed value of $\alpha$ = 0.5 for both viscous and non-viscous cases. Value of $\xi$ used is $10^6$ Pa sec. Parameters used for $\Lambda$CDM are $\Omega_m$ = 0.31, $\Omega_\Lambda$ = 0.69. There are two insets in the figure. Inset (a) shows the magnified portion of the plots at lower values of redshift, from $z = 1.4$ to $z=1.6$. Inset (b) shows the magnified portion of the plots at higher values of redshift, from $z = 4.5$ to $z = 5$.}\label{fig:DL}
\end{figure*}

In (\autoref{fig:DL}), the ratio of luminosity distance $D_L$ to the present Hubble length $D_H$ is plotted with respect to redshift $z$ for both viscous and non-viscous cases of $\Lambda$CDM and our backreaction model. $D_L$ is calculated from $D_A$ (\autoref{fig:DA}) using the relation $D_L = (1+z)^2 D_A$. There are two insets in the figure. It can be seen here too that for low redshifts (z $<$ 0.5), the curves for our model overlap with each other and with the $\Lambda$CDM curve, but as we increase the redshift, the curves start deviating from the $\Lambda$CDM curve. Inset (a) shows the magnified portion of the plots at lower values of redshift, particularly around the value of $z$ for which the plots of $D_A/D_H$ turn around in (\autoref{fig:DA}). As in (\autoref{fig:DA}), plots for the viscous backreaction model (dashed curves) have a larger magnitude than the non-viscous case (solid curves) for about $z \lessapprox 2$. For $z \gtrapprox 2.5$, plots for the viscous case have a smaller magnitude than the non-viscous case similar to plots in (\autoref{fig:DA}). This behaviour is shown in inset (b). A point to note is that this behaviour is not observed for the $\Lambda$CDM case, where viscous case plots have a larger magnitude throughout the range of $z$ of our interest. In inset (b), purple (backreaction ($\beta = 0.8$)) and red (backreaction ($\beta = 0.7$)) plot lines overlap, which is similar to the case of (\autoref{fig:DA}) at this value of $z$. Another feature of the plots for our backreaction model can be observed at higher redshift $z$. It can be seen that the difference between the solid and dashed lines (representing the non-viscous and viscous cases, respectively) increases with larger values of $\beta$. Specifically, the difference between the red lines is smallest for the lowest values of $\beta$, while the difference between the brown lines is largest for the highest values of $\beta$.

\section{Gravitational wave amplitude }\label{sec:GWamp}

The amplitude of GW from a binary of compact objects of masses $m_{1}$ and $m_{2}$ in the early inspiral stage, where Keplerian approximations are well valid, is given by (for the cross($ \times $)-polarization) \cite{maggiore} 
\begin{equation} \label{eq:GWh} 
\begin{split}
    h_{\times } = \frac{G^{5/3}(1+z)^{5/3}}{D_{L} c^{4}}\\
    *\frac{m_{1} m_{2}}{(m_{1} + m_{2})^{1/3}}(-4 \omega^{2/3}) Sin \, 2\omega t  \, , 
\end{split}    
\end{equation} 
where $ \omega $ is the observed angular frequency of the binary of compact objects  and $D_{L}$ is the luminosity distance of the binary from the observer. For the plus ($+$)-polarization, the peak of the amplitude remains identical.
For a constant observed frequency, the redshift-dependent part in the GW amplitude is $(1+z)^{5/3}/D_{L} $. 

The amplitude of GW given in (\autoref{eq:GWh}) is derived without considering the effect of viscosity on the propagation of the GW. In Ref.\cite{Goswami}, the authors have studied the effect of the viscosity of the cosmic fluid, particularly dark matter, on the GW amplitude and have estimated its effect on the GW amplitude. Assessing the impact of viscosity, after travelling a proper distance $L = ar$, the GW gets attenuated by the factor (see, Appendix A)\cite{Goswami}
\begin{equation}\label{eq:attenfactorL}
    \mathscr{A} = L_* e^{-\frac{\gamma}{2}L}/L, 
\end{equation}
where $L_*$ is the proper source distance for zero shear viscosity and $\gamma = 16\pi G\eta$, where $\eta$ is the coefficient of shear viscosity for the region through which GW is propagating. The attenuation factor in terms of luminosity distance $D_L$ is given by :
\begin{equation}\label{eq:attenfactor2}
\mathscr{A} =  \frac{D_{L_*}}{D_L} e^{-\frac{\gamma}{2(1+z)^2}D_L}, 
\end{equation}
 where $D_L = (1+z)^2 L$. $D_{L_*}$ is the luminosity distance of the source, for zero shear viscosity. Therefore, the total redshift-dependent part of the attenuated GW amplitude (let's represent this quantity by $F(z)$) is given by
\begin{equation}\label{eq:F(z)}
\begin{split}
    F(z) = \frac{(1+z)^{5/3}}{D_{L_*}}\mathscr{A}\\ = \frac{(1+z)^{5/3}}{D_{L_*}} \frac{D_{L_*}}{D_L} \, e^{-\frac{\gamma}{2(1+z)^2}D_L}\\ = \frac{(1+z)^{5/3}}{D_L} \, e^{-\frac{\gamma}{2(1+z)^2}D_L} \, .
\end{split}
\end{equation}
For non viscous case ($\eta = 0$),
\begin{equation}\label{eq:F(z)nonviscous}
    F(z,\eta = 0) = \frac{(1+z)^{5/3}}{D_L}
\end{equation}

\begin{figure*}
\includegraphics[scale = 0.8]{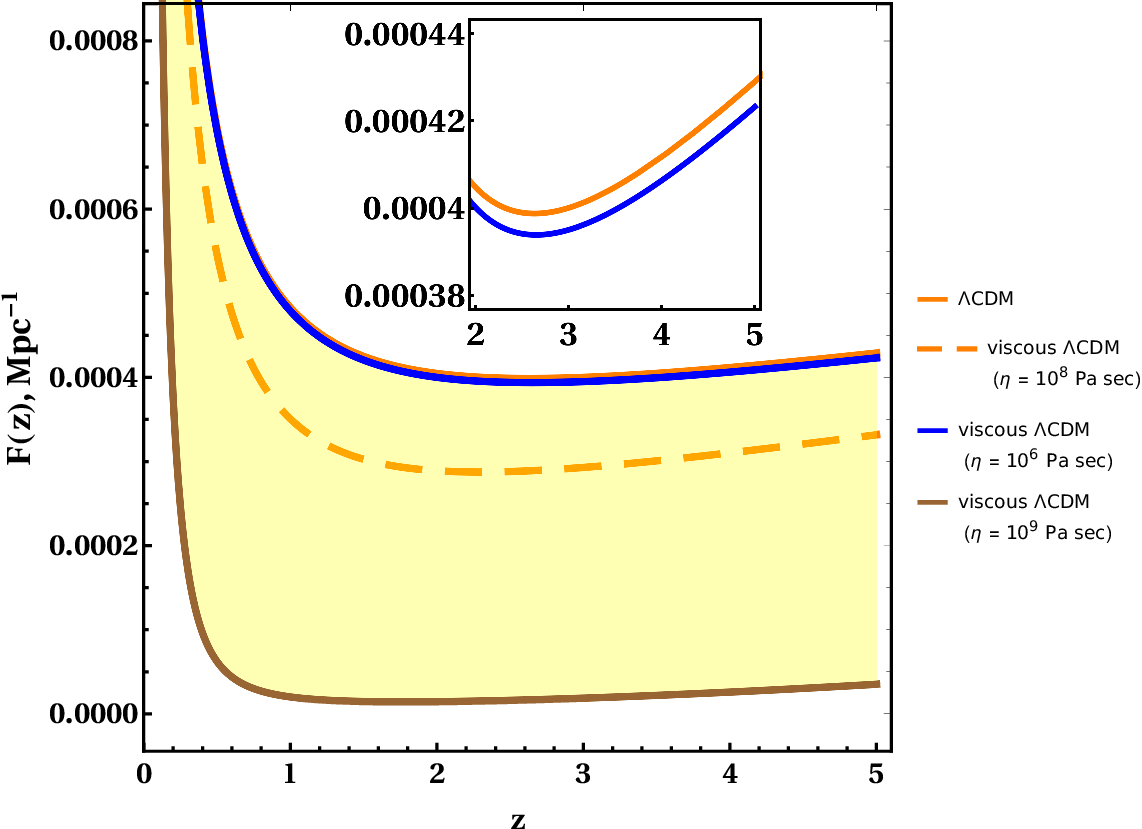}
\caption{Plot of F(z) for the $\Lambda$CDM model for both viscous and non-viscous cases. Value of $\xi$ used is $10^6$ Pa sec. Plots for viscous cases are for the applicable range of values of $\eta$  between $\sim 10^6$ Pa sec and $\sim 10^9$ Pa sec. Dashed orange plot represents the favored value of $\eta$. Parameters used for $\Lambda$CDM are $\Omega_m$ = 0.31, $\Omega_\Lambda$ = 0.69. The inset shows the magnified portion of the plot for the $\Lambda$CDM model and for our backreaction model ($\eta$ = $10^6$ Pa sec).} \label{fig:LCDMvisrange}
\end{figure*}

The luminosity distance $D_L$ is calculated from angular diameter distance, $D_A$ as $D_L = (1+z)^2 D_A$. Since, $D_L = (1+z)^2 L$,  for the flat FLRW spacetime, $D_A$ gives a good measure of the proper distance $L$. $D_A$  for flat spacetime is given by
\begin{equation}\label{eq:DA}
    D_A = \frac{D_C}{(1+z)} = \frac{D_H}{(1+z)}\int_0^z \frac{dz'}{E(z')}
\end{equation}
where $D_C$ is the comoving distance which is given as $D_H\int_0^z \frac{dz'}{E(z')}$ where $D_H$ is the Hubble distance (= $\frac{c}{H_0}$, where $H_0 =$ 100 $h$ km $s^{-1}$ $Mpc^{-1}$ $\simeq$ 0.07 $Gyr^{-1}$, with h = 0.7.) and $E(z)_{\Lambda CDM}\equiv \sqrt{\Omega_M(1+z)^3+\Omega_\Lambda}$ for the $\Lambda$CDM model (non-viscous case) (throughout this work, parameters used for $\Lambda$CDM are $\Omega_m$ = 0.31, $\Omega_\Lambda$ = 0.69. ) and $E(z)_{v\Lambda CDM}\equiv \sqrt{\Omega_v+\Omega_\Lambda}$ for the viscous $\Lambda$CDM model, where $\Omega_v$ is the fractional density for viscous matter which has a contribution from the bulk viscosity $\xi$ (\autoref{subsec:vLCDM}). Hence, now, for viscous $\Lambda$CDM case $(v\Lambda CDM)$,
\begin{equation}\label{eq:F(z)1}
    F(z)_{v\Lambda CDM}  = \frac{(1+z)^{5/3}}{D_{L_v}} \, e^{-\frac{\gamma}{2(1+z)^2}D_{L_v}} \, .
\end{equation}
where $D_{L_v}$ is the luminosity distance in the presence of viscosity, i.e. it is calculated using $E(z)_{v\Lambda CDM}$. For non viscous $\Lambda$CDM case,
\begin{equation}\label{eq:F(z)nonviscous1}
    F(z)_{\Lambda CDM} = \frac{(1+z)^{5/3}}{D_{L_*}}
\end{equation}
where $D_{L_*}$ is the source luminosity distance in the absence of viscosity, i.e. it is calculated using $E(z)_{\Lambda CDM}$.  

The value of the shear viscosity parameter $\eta$ has been estimated in
earlier works to lie within the range $\sim 10^6$ Pa sec and $\sim 10^9$ Pa sec \cite{Goswami, brevik, viscosity_lepton}. Gravitational wave observation data from LIGO has been used to put constraints on the value of $\eta$ \cite{Goswami}. Ref. \cite{viscosity_lepton} used the relativistic Boltzmann equation to examine the theory of viscosities at the time of neutrino decoupling.  The analysis of \cite{viscosity_lepton} to calculate  the present-day value of $\eta$ was subsequently employed in \cite{brevik},  to relate the value of $\eta$ at the time of neutrino decoupling with the
present value using a scaling relation, leading to the value of $10^8$ Pa sec.
Our present analysis uses the above value of $10^8$ Pa sec as the most favored
one.

In (\autoref{fig:LCDMvisrange}), we plot the overall redshift-dependent part, $F(z)$ (both $F(z)_{v\Lambda CDM}$ $\&$ $F(z)_{\Lambda CDM}$), given in (\autoref{eq:F(z)1} $\&$ \autoref{eq:F(z)nonviscous1}) w.r.t. redshift $z$, for the range of values of $\eta$  between $\sim 10^6$ Pa sec and $\sim 10^9$ Pa sec. The plots for viscous cases in (\autoref{fig:LCDMvisrange}) have contributions from both bulk viscosity $\xi$ (through $D_{L_v}$) and shear viscosity $\eta$. It was argued in Ref.~\cite{Goswami} that since $\xi$ only couples to scalar perturbations,   it doesn't play a role in the attenuation of GWs, and only $\eta$ affects the GW amplitude. However, from our analysis, it is clear that both $\xi$ and $\eta$ affect the GW amplitude, with the role of the former entering
through the modified background dynamics due to bulk viscosity.

The expressions given in (\autoref{eq:F(z)1} $\&$ \autoref{eq:F(z)nonviscous1}) are valid for a space with homogeneous mass distribution. To examine the variation of the redshift-dependent part of GW amplitude for our model, we have to modify the expressions accordingly. For our backreaction model, the relation $D_L = (1+z)^2 D_A$ is valid too, but in this case $D_A$ is calculated using the covariant scheme (\autoref{eq:covariantsch1}, \autoref{eq:covariantsch2}). Our model represents a space with inhomogeneous mass distribution, where there are two types of regions, viz. over-dense and under-dense, and viscosity is only associated with the over-dense region, as the under-dense region is assumed to be empty. 

Hence, in the  exponential term of the F(z) in (\autoref{eq:F(z)}),  $e^{-\frac{\gamma}{2(1+z)^2}D_L}$, for our model,  $D_{L_{o}}$ would replace $D_L$, where $D_{L_{o}}$ is the luminosity distance traversed by the GW through the over-dense region only. So, the  exponential term of the F(z) in (\autoref{eq:F(z)}) for our model is now given by $e^{-\frac{\gamma}{2(1+z)^2}D_{L_{o}}}$. 
In the $\Lambda$CDM model, incorporating viscosity results in an attenuation factor with $D_L$ in the exponential factor, where $D_L$ is the total luminosity distance traversed by the GW. Viscosity is not distributed through the entire path of the GW but is concentrated only in some regions. Using $D_L$ in the $\Lambda$CDM model results in a larger deviation between the attenuated and unattenuated cases in the $\Lambda$CDM model, compared to our model.

 An important consideration for the propagation of EM waves for a  model like ours (inhomogeneous 2- domain model) is that the ratio of distances travelled by EM waves through the two regions is equal to the ratio of proper volumes of the two regions \cite{Koksbang_2019}. It is clear that this also holds for GWs. This condition gives,
\begin{equation}\label{eq:ratio of DL}
    \frac{D_{L_o}}{D_{L_u}} = \left(\frac{a_o}{a_u}\right)^3  \, ,
\end{equation}
where $D_{L_o}$ and $D_{L_u}$ are the luminosity distances traversed by the GW through the over-dense and under-dense regions, respectively. 
If the total luminosity distance travelled by the GW in our model is $D_{L_{v2d}}$ (v2d stands for viscous 2-domain inhomogeneous model, this $D_{L_{v2d}}$ is calculated for the viscous case of our model (\autoref{subsubsec:viscous}) using $D_A$ calculated from the covariant scheme (\autoref{eq:covariantsch1}, \autoref{eq:covariantsch2})), then $D_{L_{v2d}} = D_{L_o} + D_{L_u}$, which gives, 
\begin{equation*}
     D_{L_o}\left(1+\frac{D_{L_u}}{D_{L_o}}\right) = D_{L_o}\left(1+\left(\frac{a_u}{a_o}\right)^3\right) = D_{L_{v2d}} \, , 
\end{equation*}
or, 
\begin{equation} \label{eq:DLo}
    D_{L_o} = \frac{D_{L_{v2d}}}{\left(1+\left(\frac{a_u}{a_o}\right)^3\right)}  \, . 
\end{equation}
Therefore, the exponential term in the total redshift dependent part of attenuated 
 GW amplitude for our model now becomes,
\begin{equation} \label{eq:attenfactor3}
   \mathcal{E} = e^{-\frac{\gamma}{2(1+z)^2}\frac{D_{L_{v2d}}}{\left(1+\left(\frac{a_u}{a_o}\right)^3\right)}}, 
\end{equation}
and the total redshift-dependent part of the attenuated GW amplitude for our viscous model is given by,
\begin{equation}\label{eq:F(z)2}
\begin{split}
    F(z)_{v2d} = \frac{(1+z)^{5/3}}{D_{L_{v2d}}} \mathcal{E} \\
    = \frac{(1+z)^{5/3}}{D_{L_{v2d}}}  e^{ -\frac{\gamma}{2(1+z)^2}\frac{D_{L_{v2d}}}{\left(1+\left(\frac{a_u}{a_o}\right)^3\right)}}  \, .
\end{split} 
\end{equation}
The redshift-dependent part of the GW amplitude for our non-viscous model (\autoref{subsubsec:nonviscous}) is given by,
\begin{equation}\label{eq:F(z)nonviscous2}
   F(z, \xi = 0, \eta=0)_{2d} = \frac{(1+z)^{5/3}}{{D_{L_{*2d}}}}  \, . 
\end{equation}
where 2d stands for non-viscous 2-domain inhomogeneous model and $D_{L_{*2d}}$ is the source luminosity distance for the case of our non-viscous 2-domain inhomogeneous model.
Therefore, for our model, the deviation of the redshift-dependent part of GW amplitude due to viscous-attenuation, can be written as, 
\begin{dmath}\label{eq:deviation}
    F(z, \xi = 0, \eta=0)_{2d}-F(z)_{v2d} 
    = (1+z)^{5/3} \left\lbrace \frac{1}{{D_{L_{2d}}}} -  \frac{e^{-\frac{\gamma}{2(1+z)^2}\frac{D_{L_{v2d}}}{\left(1+\left(\frac{a_u}{a_o}\right)^3\right)}}}{D_{L_{v2d}}} \right\rbrace  \, . 
\end{dmath}
\begin{figure*}
    \includegraphics[scale = 0.8]{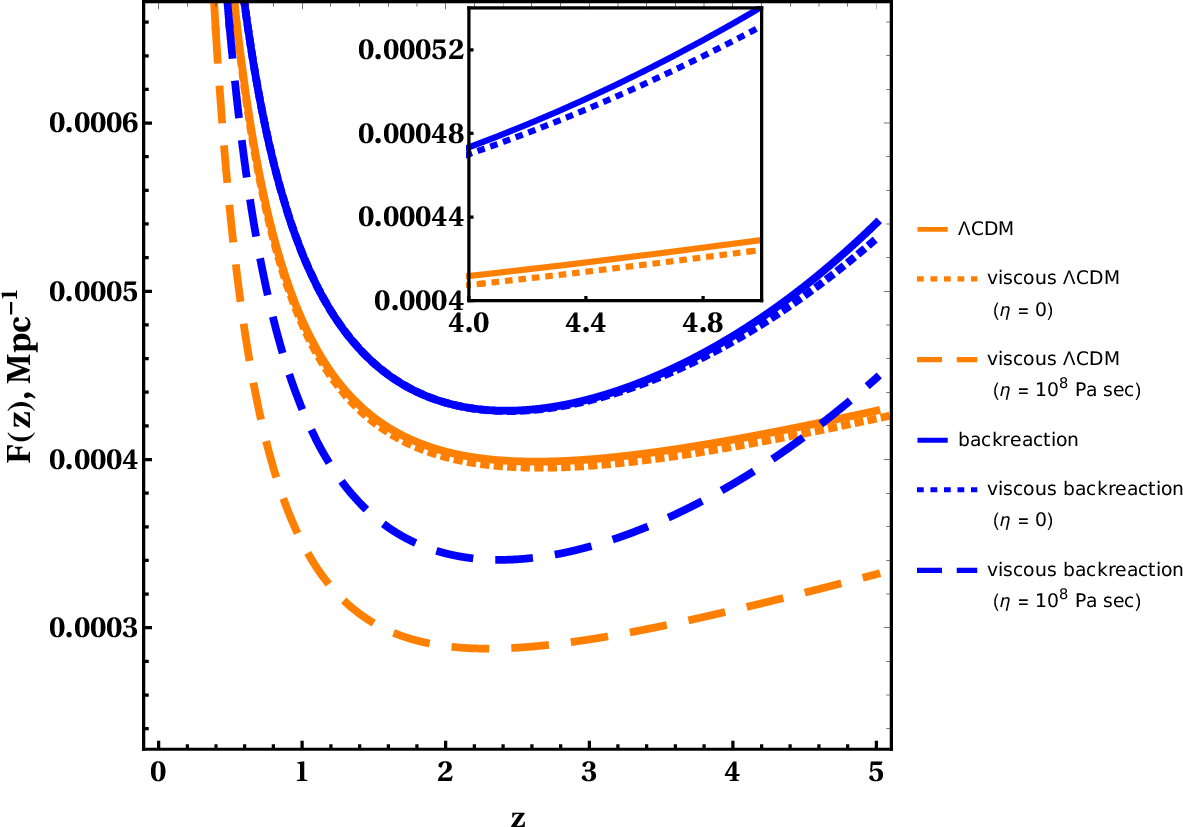}
    \caption{Plot of $F(z)$ vs z for $\Lambda$CDM model and for our model for $(\alpha,\beta)$ = (0.67,1). The solid curves are for non-viscous cases. The dotted curves are for those cases in which only $\xi$ has been included. The dashed curves have contributions from both $\xi$ and $\eta$. Value of $\xi$ used is $10^6$ Pa sec. Parameters used for $\Lambda$CDM are $\Omega_m$ = 0.31, $\Omega_\Lambda$ = 0.69. Inset shows the magnified portion of the solid curve and the dotted curve for the two models to illustrate the difference between these two curves. }
    \label{fig:GWatten3}
\end{figure*}

To summarize, the redshift dependent part of GW for a homogeneous and non-viscous spacetime ($\Lambda$CDM model) is given by (\autoref{eq:F(z)nonviscous1}). Now, if we introduce matter distribution inhomogeneities, then F(z) gets modified due to modification of the redshift-distance relations, and is now given by (\autoref{eq:F(z)nonviscous2}). It can be seen that,
\begin{equation}\label{eq:2dLCDM}
    F(z, \xi = 0, \eta=0)_{2d} = F(z)_{\Lambda CDM}\times \frac{D_{L_*}}{D_{L_{*2d}}}
\end{equation}
Therefore, the effect of inclusion of matter distribution inhomogeneities on redshift dependent part of GW amplitude is equivalent to multiplication by the factor $\frac{D_{L_*}}{D_{L_{*2d}}}$. 
\par On further introduction of viscosity in the analysis, GW in our backreaction model gets attenuated by the attenuation factor (using (\autoref{eq:attenfactor3})),
\begin{equation}\label{eq:finalA}
\begin{split}
    \mathfrak{A} = \frac{D_{L_{*2d}}}{D_{L_{v2d}}}\mathcal{E}\\
    = \frac{D_{L_{*2d}}}{D_{L_{v2d}}}e^{-\frac{\gamma}{2(1+z)^2}\frac{D_{L_{v2d}}}{\left(1+\left(\frac{a_u}{a_o}\right)^3\right)}}
\end{split}
\end{equation}
where $D_{L_{*2d}}$ is the luminosity distance for the case of our non-viscous 2-domain inhomogeneous model, calculated using the covariant scheme. 

\par Thus, the total redshift dependent part of GW in the presence of viscous inhomogeneities becomes (using (\autoref{eq:F(z)2}),(\autoref{eq:F(z)nonviscous2}), (\autoref{eq:2dLCDM}) $\&$ (\autoref{eq:finalA})),
\begin{equation}
    \begin{split}
        F(z)_{v2d} = F(z)_{\Lambda CDM}\times \frac{D_{L_*}}{D_{L_{*2d}}}\times \mathfrak{A}\\
        = F(z, \xi = 0, \eta=0)_{2d} \times \mathfrak{A}\\
        = \frac{(1+z)^{5/3}}{{D_{L_{*2d}}}} \times \frac{D_{L_{*2d}}}{D_{L_{v2d}}}\mathcal{E}\\
        =  \frac{(1+z)^{5/3}}{D_{L_{v2d}}}\mathcal{E} \\
        = \frac{(1+z)^{5/3}}{D_{L_{v2d}}}  e^{ -\frac{\gamma}{2(1+z)^2}\frac{D_{L_{v2d}}}{\left(1+\left(\frac{a_u}{a_o}\right)^3\right)}}
    \end{split}
\end{equation}

\par In (\autoref{fig:GWatten3}), we plot the redshift dependent part of the GW amplitude, $F(z)$ vs z for the $\Lambda$CDM model and for our model for model parameter $(\alpha,\beta) = (0.67,1)$. There are 3 curves for each model. The solid curve represents the non-viscous case; the dotted curve represents the case in which only the bulk viscosity has been included in the analysis, and the dashed curve is the case with both bulk viscosity and shear viscosity in the analysis. As can be seen from (\autoref{fig:GWatten3}), even if we just consider bulk viscosity then also there is deviation of the redshift dependent part of GW amplitude (dotted curve) with respect to the non viscous case (solid curve). This is because redshift dependent part of GW amplitude consists of $D_L$ and $\xi$ affects this $D_L$ via the quantity $E(z)$ (\autoref{eq:DA}). On further inclusion of $\eta$ in the analysis, redshift dependent part of GW amplitude gets attenuated (dashed curve). In this case, $F(z)$ for viscous $\Lambda$CDM model is given by (\autoref{eq:F(z)1}), for non viscous $\Lambda$CDM model by (\autoref{eq:F(z)nonviscous1}), for our non viscous model with inhomogeneities by (\autoref{eq:F(z)nonviscous2}) and for our viscous model with inhomogeneities, it is given by (\autoref{eq:F(z)2}).

\begin{figure*}
    \includegraphics[scale = 0.6]{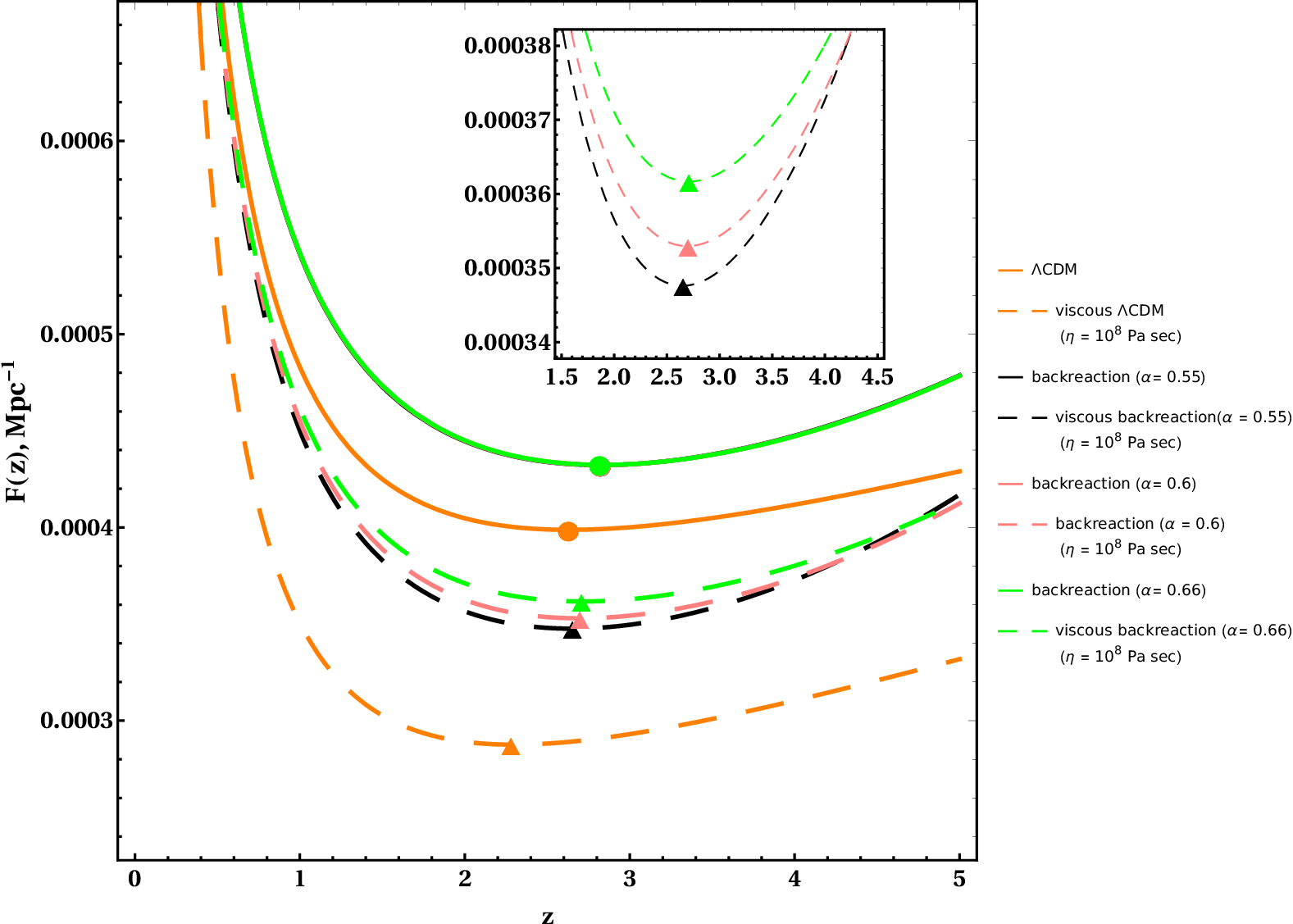}
    \caption{Plot of $F(z)$ vs z for $\Lambda$CDM model and for our model for different combinations of $(\alpha,\beta)$ while keeping $\beta$ = 0.8 as constant. The solid curves are for non-viscous cases. The viscous cases represented by dashed curves have contributions from both $\xi$ and $\eta$. Value of $\xi$ used is $10^6$ Pa sec. Parameters used for $\Lambda$CDM are $\Omega_m$ = 0.31, $\Omega_\Lambda$ = 0.69. Circular points represent the minima for the solid (non-viscous)curves while triangular points represent the minima positions for the dashed (viscous) curves. Inset shows the magnified portion of the three dashed curves of our backreaction model to illustrate the difference between these curves.}
    \label{fig:GWattenconst}
\end{figure*}

In (\autoref{fig:GWattenconst}), we plot the redshift dependent part of the GW amplitude, $F(z)$ vs z for the $\Lambda$CDM model and for our model for different combinations of model parameter $(\alpha,\beta)$. Here, we have kept $\beta$ constant $=0.8$, and we  vary the value of $\alpha$. The solid curves represent the non-viscous cases, and the dashed curves represent the viscous case with contributions from both $\xi$ and $\eta$. The redshift dependent part of GW amplitude, i.e., the quantity $(1+z)^{5/3}/D_L$ in the case of $\Lambda$CDM, has a minimum at $z_{min}\simeq 2.63$ \cite{Rosado}.  In (\autoref{fig:GWattenconst}) we display the minima points of the various curves. Circular points represent the minima for the solid (non-viscous) curves while triangular points represent the minima positions for the dashed (viscous) curves.  We observe that the minima for the v$\Lambda$CDM curve  deviates significantly from the $\Lambda$CDM case. For our backreaction model (both viscous and non viscous), minima points  deviate significantly from the $\Lambda$CDM case. It can be seen from the figure that for the non-viscous cases of our model, all three plot lines and their minima points for different values of $\alpha$ overlap. $F(z)$ for the non-viscous cases is given by (\autoref{eq:F(z)nonviscous2}). Since the value of $(1+z)^{5/3}$ is the same for all $(\alpha,\beta)$, the only quantity varying with changing $(\alpha,\beta)$ is $D_L$. The volume fraction of the overdense region at the present time is taken as 0.09, and the range of variation of $\alpha$, which governs the evolution of the overdense region, is from $0.5 - 0.67$. Since the overdense volume fraction is so small; therefore variation of $\alpha$ over this small range (0.5 - 0.67) doesn't have much effect, and hence, the plot lines for the non-viscous cases  overlap in (\autoref{fig:GWattenconst}).

From (\autoref{fig:GWattenconst}), it can be seen that there is substantial attenuation of the redshift-dependent part of the GW amplitude due to viscosity. $F(z)$ for viscous cases is given by (\autoref{eq:F(z)2}). As compared to the non-viscous case (\autoref{eq:F(z)nonviscous2}), there are now additional terms. From (\autoref{eq:aomod2}) and (\autoref{eq:co_modparam}), the scale factor of the overdense region is given by, 
\begin{equation}
    a_o = \left(\frac{t}{t_0}\right)^{\alpha} = \left(\frac{t}{13.8}\right)^{\alpha} \label{eq:ao2}
\end{equation}
where $t_0 = 13.8$ Gyr. For a given value of t (t $<$ $t_0$), as we increase $\alpha$, $a_o$ decreases. It can be seen in (\autoref{fig:GWattenconst}), as we increase the value of $\alpha$ keeping $\beta$ constant, the redshift dependent part of GW, $F(z)$, for the viscous case, gets smaller in magnitude. Physically, this can be explained from the fact that from (\autoref{eq:aD}) for an overdense region, $a_o$ can be defined as
\begin{equation}\label{eq:ao3}
    a_o (t) := \left(\frac{V_o(t)}{V_{o_0}}\right)^{1/3} \, , 
\end{equation}
where $V_{o_0}$ is the volume of the overdense region at the present time which we have fixed for all values of $(\alpha,\beta)$ to be 0.09 by fraction of the total volume. As $a_o$ decreases, the volume of the overdense domain decreases. Hence, the distance travelled by the GW through the overdense region will be less, leading to less attenuation suffered by the GW. GW which travels the largest distance through the overdense region will suffer the most attenuation. This distance travelled by the GW is directly proportional to the scale factor of the overdense region. In (\autoref{fig:GWattenconst}), the green dashed curve has the largest value of $\alpha$ (hence the smallest value of $a_o$), corresponding to the largest 
amplitude among the viscous cases since it has suffered the least attenuation. 
In contrast, the black dashed curve has the smallest value of $\alpha$ (hence the largest value of $a_o$), 
corresponding to the smallest amplitude among the viscous cases since it has suffered the most attenuation. The above feature is not observed for the  non-viscous cases (solid curves) as the overdense region is non-viscous; hence, there is no viscous-attenuation.

\section{Conclusions}\label{sec:conclusion}

In this work, we have studied the propagation of GWs from compact binary sources through a viscous inhomogeneous Universe governed by a model  based on the averaging procedure for scalars in Buchert's backreaction framework \cite{Buchert, Weigand_et_al}. Dynamics under this model lead to a modification of the redshift-versus-distance relation from that for the $\Lambda$CDM model. The extent of variation depends on the combination of the model parameters. In the present work, we have considered viscosity to be present in the matter content within the over-dense regions of inhomogeneous spacetime described by our model, which causes the attenuation of GW amplitude when the GW passes through those regions of spacetime. We have incorporated the viscous attenuation of GW amplitude within our model of inhomogeneous spacetime and have derived an expression for the resultant redshift-dependent part of the GW amplitude.

In the $\Lambda$CDM model, incorporating viscosity results in a GW attenuation factor with $D_L$ in the exponent, where $D_L$ is the total luminosity distance traversed by the GW. It is worth noting that in the real Universe, viscosity 
resulting from dark matter interactions is not distributed uniformly through the entire path of the GW but is concentrated only in some regions. Therefore, in our model, we have used $D_{L_o}$ (luminosity distance of the overdense region) instead of $D_L$. Using $D_L$ results in a more significant deviation between the attenuated and unattenuated cases for the $\Lambda$CDM model, compared to using $D_{L_o}$ for our model. It has been argued earlier ~\cite{Goswami} that 
since bulk viscosity   couples only to scalar perturbations,  it doesn't play a role in the attenuation of GWs. However, as shown here, bulk
viscosity indirectly impacts the GW amplitude through its effect on the
luminosity distance. Moreover, the effect of shear viscosity on GW attenuation
is clearly demonstrated in the backreaction model due to inhomogeneities.

Our analysis demonstrates a substantial deviation in the redshift-dependent part of the GW amplitude due to the inclusion of viscous attenuation, compared to the case when viscosity is considered negligible or absent within the model of inhomogeneous spacetime. We have further shown that the rate
of expansion of the overdense region (characterized by the parameter $\alpha$, which governs the time evolution of the scale factor of the overdense region) plays a vital role in the magnitude of attenuation. It may be 
emphasized that consideration of the effect of viscosity on GW observables for compact binary sources is significant in the context of local inhomogeneities
in the Universe. 

To summarize, we would again like to highlight the importance of considering realistic background effects in the study of GW propagation, since using incorrect background dynamics for analyzing GW data from detectors could result in incorrect inferences about GW sources. Towards this end we have taken into account  these two aspects of inhomogeneities and attenuation of GW due to viscosity together in our present work. Our analysis leads to several interesting features in the red-shift dependent part of the gravitational wave amplitude. Additionally, the role of bulk viscosity  is a new feature that is also brought out in the form on its indirect contribution towards GW attenuation through modification of the background dynamics.Our analysis paves the way for obtaining more precise estimations of GW observables and bounds on the viscosity parameters of dark matter in future work involving more realistic backreaction models and data-analysis techniques in GW astronomy.

\vspace{0.2cm}
\section*{Acknowledgements}
SSP thanks the Council of Scientific and Industrial Research (CSIR), Govt. of India, for funding through the CSIR-JRF-NET fellowship.\\
This version of the article has been accepted for publication, after peer review (when applicable) but is not the Version of Record and does not reflect post-acceptance improvements, or any corrections. The Version of Record is available online at: \hyperlink{https://doi.org/10.1140/epjc/s10052-023-11605-9}{https://doi.org/10.1140/epjc/s10052-023-11605-9}.

\section*{Declarations}
\begin{itemize}
\item  The authors did not receive support from any organization for the submitted work.
\item The authors have no competing interests to declare that are relevant to the content of this article.
\end{itemize}

\begin{appendices}

\section{GWs in the presence of viscosity}\label{appendixA}

GWs are affected by the viscosity of the propagating medium 
\cite{Hawking,Anile1978,PRASANNA1999120,Esposito,Madore1973,Goswami,brevik}. It was shown \cite{PRASANNA1999120} that only the coefficient of shear viscosity has any influence on the attenuation of GW.
\par Here, we discuss some essential steps in the
derivation of the attenuation factor of a GW propagating through a viscous fluid in an FRW Universe. The same relation of the attenuation factor is also valid for our backreaction models, incorporating the redshift-distance relation given by the covariant scheme.  We use the mechanism employed by \cite{Goswami}. The general form of the energy-momentum tensor for a non-ideal fluid is given by \cite{Goswami, Weinberg},
\begin{equation}\label{eq:Tmunu}
    T_{\mu\nu} = (\rho+p)u_{\mu}u_{\nu} + pg_{\mu\nu} - 2\eta\sigma_{\mu\nu} - \xi\theta\Delta_{\mu\nu}
\end{equation}
where $\rho$ is the density of the fluid, p is the pressure of the fluid, $u_{\mu}$ is the fluid four velocity, $g_{\mu\nu}$ is the metric tensor, $\eta$ is the coefficient of shear viscosity, $\sigma_{\mu\nu}$ is the shear, $\xi$ is the coefficient of bulk viscosity, $\theta$ is the volume expansion of the fluid and $\Delta_{\mu\nu}$ is the projection tensor on the subspace normal to $u_{\mu}$ and it is given by the relation, $\Delta_{\mu\nu} = g_{\mu\nu}+u_{\mu}u_{\nu}$. Throughout, $\hbar = c = 1$.
\par Next, we consider tensor perturbations in the background FRW metric,
\begin{equation}\label{eq:GWmetric}
    ds^2 = -dt^2+a^2(t)[\delta_{ij}+h_{ij}]dx^idx^j,
\end{equation}
The tensor perturbations are considered in the transverse and traceless gauge, $\partial^ih_{ij} = h^i_i = 0$.
\par The total four-velocity is given by $u_{\mu} = u_{\mu}^{(0)} + \partial u_\mu$. Normalizing the four-velocity, considering only up to first-order terms in the metric and velocity perturbations, and  going to the rest frame of the fluid, the velocity perturbations $\partial u^\mu$ vanish. In the rest frame of the fluid, we have,
\begin{equation}
    \theta = 3H
\end{equation}
\begin{equation}
    \sigma_{ij} = \frac{1}{2}a^2\dot{h}_{ij},
\end{equation}
where H is the Hubble parameter and dot denotes derivative with respect to cosmic time t.
\par Einstein's equation in zeroth order in $h_{ij}$, $G_{ij} = 8\pi G T_{ij}$ and in first order in $h_{ij}$, $\partial G_{ij} = 8\pi G T_{ij}$ gives us the wave equation for GWs in a viscous fluid,
\begin{equation}\label{eq:hij}
    \ddot{h}_{ij} + (3H + 16\pi G\eta)\dot{h}_{ij} - \frac{\nabla^2}{a^2}h_{ij} = 0
\end{equation}
$\xi$ doesn't come into (\autoref{eq:hij}) as it only couples to scalar perturbations.
\par Next, performing the Fourier transform of (\autoref{eq:hij}) and defining $h_{ij}$ as $\mu_{ij}/a$ one gets,
\begin{equation}
\begin{split}\label{eq:muij}
    \ddot{\mu}_{ij} + (H +16\pi G\eta)\dot{\mu}_{ij}\\
    + \left(\frac{k^2}{a^2}-\frac{\ddot{a}}{a} - H^2 - 16\pi G\eta H\right)\mu_{ij} = 0
\end{split}
\end{equation}
\par Defining conformal time $\tau$ as $dt = ad\tau$ and using it in (\autoref{eq:muij}) leads to
\begin{equation}
    \begin{split}\label{eq:muij2}
        \mu''_{ij}+ 16\pi G\eta a \mu'_{ij}\\
        + \left(k^2 - \frac{a''}{a}-16\pi G\eta a \mathcal{H} \right)\mu_{ij} = 0
    \end{split}
\end{equation}
where $'$ denotes derivatives with respect to $\tau$. On sub-horizon scales $k^2>>\frac{a'}{a}$, (\autoref{eq:muij2}) reduces to,
\begin{equation}
    \mu''_{ij} + 16\pi G\eta a \mu'_{ij} + k^2\mu_{ij} = 0.
\end{equation}
Let $A_{\times,+} = r\mu_{ij}$ represent the amplitude of the two polarization modes $\times$ and $+$ of the radial component of the wave. Then, at large distances from the source, $A$ satisfies the following 1-D wave equation,
\begin{equation}\label{eq:Adotdot}
    \ddot{A} + \beta a \dot{A} + k^2A = 0
\end{equation}
where $\beta \equiv 16\pi G \eta$. Assuming the solution of (\autoref{eq:Adotdot}) is of the form,
\begin{equation}\label{eq:Asol}
    A(\tau,\omega)= \Tilde{A}(\omega)e^{ikr-\int i\omega d\tau}.
\end{equation}
and substituting (\autoref{eq:Asol}) in (\autoref{eq:Adotdot}), gives us the dispersion relation,
\begin{equation}\label{eq:disp}
    -\omega^2 - i\beta a\omega+k^2 = 0
\end{equation}
Separating the real and imaginary part of k, $k = k_R+ik_I$, (\autoref{eq:disp}) in conjugation with the weak damping approximation $\beta<<\omega$, and keeping only the leading order terms gives us,
\begin{equation}
    \begin{split}
        k_R = \omega\\
        k_I = \frac{\beta a}{2}
    \end{split}
\end{equation}
Presence of imaginary part of k, $k_I$ results in attenuation of the wave. (\autoref{eq:Asol}) now becomes,
\begin{equation}
    A(\tau,\omega)= \Tilde{A}(\omega)e^{ik_Rr-\int i\omega d\tau}\times e^{-k_Ir}.
\end{equation}
The strain $h_{ij}$ of the GW in cosmic time t now becomes,
\begin{equation}
    h_{ij} = \frac{\Tilde{A}(\omega(t))}{L_0}e^{ik_Rr-\int i\omega_p dt}\times \frac{L_0 e^{-\frac{\beta}{2}L}}{L}
\end{equation}
where $L = ar$ is the source distance, $L_0$ is the source distance for zero shear viscosity and $\omega_p = \frac{\omega}{a}$ is the physical angular frequency. Therefore, the attenuation factor is given by $\frac{L_0 e^{-\frac{\beta}{2}L}}{L}$ where $\beta \equiv 16\pi G\eta$.




\end{appendices}


\bibliography{sn-bibliography}

\begin{thebibliography}{10}
\providecommand{\url}[1]{{#1}}
\providecommand{\urlprefix}{URL }
\providecommand{\doi}[1]{\url{https://doi.org/#1}}
\bibcommenthead

\bibitem{Einstein1}
A.~Einstein, \emph{Näherungsweise Integration der Feldgleichungen der
  Gravitation} (John Wiley $\&$ Sons, Ltd, 2005), pp. 99--108.
\newblock \doi{https://doi.org/10.1002/3527608958.ch7}.
\newblock
  \urlprefix\url{https://onlinelibrary.wiley.com/doi/abs/10.1002/3527608958.ch7}

\bibitem{Einstein2}
A.~{Einstein}, {{\"U}ber Gravitationswellen}.
\newblock Sitzungsberichte der K{\"o}niglich Preu{\ss}ischen Akademie der
  Wissenschaften (Berlin pp. 154--167 (1918)

\bibitem{Abbott_et_al}
B.P. Abbott~\textit{et al}, Observation of gravitational waves from a binary
  black hole merger.
\newblock Phys. Rev. Lett. \textbf{116}, 061,102 (2016).
\newblock \doi{10.1103/PhysRevLett.116.061102}.
\newblock
  \urlprefix\url{https://link.aps.org/doi/10.1103/PhysRevLett.116.061102}

\bibitem{Ligo-virgo}
B.P. Abbott~\textit{et al}, Gw151226: Observation of gravitational waves from a
  22-solar-mass binary black hole coalescence.
\newblock Phys. Rev. Lett. \textbf{116}, 241,103 (2016).
\newblock \doi{10.1103/PhysRevLett.116.241103}.
\newblock
  \urlprefix\url{https://link.aps.org/doi/10.1103/PhysRevLett.116.241103}

\bibitem{Ligo-virgo2}
B.P. Abbott~\textit{et al}, Gw170104: Observation of a 50-solar-mass binary
  black hole coalescence at redshift 0.2.
\newblock Phys. Rev. Lett. \textbf{118}, 221,101 (2017).
\newblock \doi{10.1103/PhysRevLett.118.221101}.
\newblock
  \urlprefix\url{https://link.aps.org/doi/10.1103/PhysRevLett.118.221101}

\bibitem{Ligo-virgo3}
B.P. Abbott~\textit{et al}, Gw170814: A three-detector observation of
  gravitational waves from a binary black hole coalescence.
\newblock Phys. Rev. Lett. \textbf{119}, 141,101 (2017).
\newblock \doi{10.1103/PhysRevLett.119.141101}.
\newblock
  \urlprefix\url{https://link.aps.org/doi/10.1103/PhysRevLett.119.141101}

\bibitem{Ligo-virgo4}
B.P. Abbott~\textit{et al}, Gw170817: Observation of gravitational waves from a
  binary neutron star inspiral.
\newblock Phys. Rev. Lett. \textbf{119}, 161,101 (2017).
\newblock \doi{10.1103/PhysRevLett.119.161101}.
\newblock
  \urlprefix\url{https://link.aps.org/doi/10.1103/PhysRevLett.119.161101}

\bibitem{Ligo-virgo5}
B.P. Abbott~\textit{et al}, {GW}170608: Observation of a 19 solar-mass binary
  black hole coalescence.
\newblock The Astrophysical Journal \textbf{851}(2), L35 (2017).
\newblock \doi{10.3847/2041-8213/aa9f0c}.
\newblock \urlprefix\url{https://doi.org/10.3847/2041-8213/aa9f0c}

\bibitem{PRASANNA1999120}
A.~Prasanna, Propagation of gravitational waves through a dispersive medium.
\newblock Physics Letters A \textbf{257}(3), 120--122 (1999).
\newblock \doi{https://doi.org/10.1016/S0375-9601(99)00313-8}.
\newblock
  \urlprefix\url{https://www.sciencedirect.com/science/article/pii/S0375960199003138}

\bibitem{Hawking}
S.W. {Hawking}, {Perturbations of an Expanding Universe}.
\newblock The Astrophysical Journal \textbf{145}, 544 (1966).
\newblock \doi{10.1086/148793}

\bibitem{Esposito}
F.P. {Esposito}, {Absorption of Gravitational Energy by a Viscous Compressible
  Fluid}.
\newblock The Astrophysical Journal \textbf{165}, 165 (1971).
\newblock \doi{10.1086/150884}

\bibitem{Madore1973}
J.~Madore, The absorption of gravitational radiation by a dissipative fluid.
\newblock Communications in Mathematical Physics \textbf{30}(4), 335--340
  (1973).
\newblock \doi{10.1007/BF01645508}.
\newblock \urlprefix\url{https://doi.org/10.1007/BF01645508}

\bibitem{Anile1978}
A.M. Anile, V.~Pirronello, High-frequency gravitational waves in a dissipative
  fluid.
\newblock Il Nuovo Cimento B (1971-1996) \textbf{48}(1), 90--101 (1978).
\newblock \doi{10.1007/BF02748651}.
\newblock \urlprefix\url{https://doi.org/10.1007/BF02748651}

\bibitem{Goswami}
G.~Goswami, G.K. Chakravarty, S.~Mohanty, A.R. Prasanna, Constraints on
  cosmological viscosity and self-interacting dark matter from gravitational
  wave observations.
\newblock Phys. Rev. D \textbf{95}, 103,509 (2017).
\newblock \doi{10.1103/PhysRevD.95.103509}.
\newblock \urlprefix\url{https://link.aps.org/doi/10.1103/PhysRevD.95.103509}

\bibitem{brevik}
I.~Brevik, S.~Nojiri, Gravitational waves in the presence of viscosity.
\newblock International Journal of Modern Physics D \textbf{28}(10), 1950,133
  (2019).
\newblock \doi{10.1142/S0218271819501335}.
\newblock \urlprefix\url{https://doi.org/10.1142/S0218271819501335}.
\newblock
  {\href{https://arxiv.org/abs/https://doi.org/10.1142/S0218271819501335}{{https://doi.org/10.1142/S0218271819501335}}}

\bibitem{Nahuel}
N.~Mirón-Granese, Relativistic viscous effects on the primordial gravitational
  waves spectrum.
\newblock Journal of Cosmology and Astroparticle Physics \textbf{2021}(06), 008
  (2021).
\newblock \doi{10.1088/1475-7516/2021/06/008}.
\newblock \urlprefix\url{https://dx.doi.org/10.1088/1475-7516/2021/06/008}

\bibitem{Moretti2020}
F.~Moretti, F.~Bombacigno, G.~Montani, Gravitational landau damping for massive
  scalar modes.
\newblock The European Physical Journal C \textbf{80}(12), 1203 (2020).
\newblock \doi{10.1140/epjc/s10052-020-08769-z}.
\newblock \urlprefix\url{https://doi.org/10.1140/epjc/s10052-020-08769-z}

\bibitem{universe7120496}
F.~Moretti, F.~Bombacigno, G.~Montani, The role of longitudinal polarizations
  in horndeski and macroscopic gravity: Introducing gravitational plasmas.
\newblock Universe \textbf{7}(12) (2021).
\newblock \doi{10.3390/universe7120496}.
\newblock \urlprefix\url{https://www.mdpi.com/2218-1997/7/12/496}

\bibitem{PADMANABHAN1987433}
T.~Padmanabhan, S.~Chitre, Viscous universes.
\newblock Physics Letters A \textbf{120}(9), 433--436 (1987).
\newblock \doi{https://doi.org/10.1016/0375-9601(87)90104-6}.
\newblock
  \urlprefix\url{https://www.sciencedirect.com/science/article/pii/0375960187901046}

\bibitem{Murphy}
G.L. Murphy, Big-bang model without singularities.
\newblock Phys. Rev. D \textbf{8}, 4231--4233 (1973).
\newblock \doi{10.1103/PhysRevD.8.4231}.
\newblock \urlprefix\url{https://link.aps.org/doi/10.1103/PhysRevD.8.4231}

\bibitem{Fabris2006}
J.C. Fabris, S.V.B. Gon{\c{c}}alves, R.d.S. Ribeiro, Bulk viscosity driving the
  acceleration of the universe.
\newblock General Relativity and Gravitation \textbf{38}(3), 495--506 (2006).
\newblock \doi{10.1007/s10714-006-0236-y}.
\newblock \urlprefix\url{https://doi.org/10.1007/s10714-006-0236-y}

\bibitem{Gagnon_2011}
J.S. Gagnon, J.~Lesgourgues, Dark goo: bulk viscosity as an alternative to dark
  energy.
\newblock Journal of Cosmology and Astroparticle Physics \textbf{2011}(09),
  026--026 (2011).
\newblock \doi{10.1088/1475-7516/2011/09/026}.
\newblock \urlprefix\url{https://doi.org/10.1088/1475-7516/2011/09/026}

\bibitem{Floerchinger}
S.~Floerchinger, N.~Tetradis, U.A. Wiedemann, Accelerating cosmological
  expansion from shear and bulk viscosity.
\newblock Phys. Rev. Lett. \textbf{114}, 091,301 (2015).
\newblock \doi{10.1103/PhysRevLett.114.091301}.
\newblock
  \urlprefix\url{https://link.aps.org/doi/10.1103/PhysRevLett.114.091301}

\bibitem{Atreya_2018}
A.~Atreya, J.R. Bhatt, A.~Mishra, Viscous self interacting dark matter and
  cosmic acceleration.
\newblock Journal of Cosmology and Astroparticle Physics \textbf{2018}(02),
  024--024 (2018).
\newblock \doi{10.1088/1475-7516/2018/02/024}.
\newblock \urlprefix\url{https://doi.org/10.1088/1475-7516/2018/02/024}

\bibitem{Mohan2017}
N.D.J. Mohan, A.~Sasidharan, T.K. Mathew, Bulk viscous matter and recent
  acceleration of the universe based on causal viscous theory.
\newblock The European Physical Journal C \textbf{77}(12), 849 (2017).
\newblock \doi{10.1140/epjc/s10052-017-5428-y}.
\newblock \urlprefix\url{https://doi.org/10.1140/epjc/s10052-017-5428-y}

\bibitem{Das2012}
S.~Das, N.~Banerjee, Can neutrino viscosity drive the late time cosmic
  acceleration?
\newblock International Journal of Theoretical Physics \textbf{51}(9),
  2771--2778 (2012).
\newblock \doi{10.1007/s10773-012-1152-4}.
\newblock \urlprefix\url{https://doi.org/10.1007/s10773-012-1152-4}

\bibitem{brevik2}
I.~Brevik, O.~Gr\o{}n, J.~de~Haro, S.D. Odintsov, E.N. Saridakis, Viscous
  cosmology for early- and late-time universe.
\newblock International Journal of Modern Physics D \textbf{26}(14), 1730,024
  (2017).
\newblock \doi{10.1142/S0218271817300245}.
\newblock \urlprefix\url{https://doi.org/10.1142/S0218271817300245}.
\newblock
  {\href{https://arxiv.org/abs/https://doi.org/10.1142/S0218271817300245}{{https://doi.org/10.1142/S0218271817300245}}}

\bibitem{Anand_2018}
S.~Anand, P.~Chaubal, A.~Mazumdar, S.~Mohanty, P.~Parashari, Bounds on neutrino
  mass in viscous cosmology.
\newblock Journal of Cosmology and Astroparticle Physics \textbf{2018}(05),
  031--031 (2018).
\newblock \doi{10.1088/1475-7516/2018/05/031}.
\newblock \urlprefix\url{https://doi.org/10.1088/1475-7516/2018/05/031}

\bibitem{Halder_2022}
A.~Halder, S.S. Pandey, A.~Majumdar, Global 21-cm brightness temperature in
  viscous dark energy models.
\newblock Journal of Cosmology and Astroparticle Physics \textbf{2022}(10), 049
  (2022).
\newblock \doi{10.1088/1475-7516/2022/10/049}.
\newblock \urlprefix\url{https://dx.doi.org/10.1088/1475-7516/2022/10/049}

\bibitem{Natwariya2020}
P.K. Natwariya, J.R. Bhatt, A.K. Pandey, Viscosity in cosmic fluids.
\newblock The European Physical Journal C \textbf{80}(8), 767 (2020).
\newblock \doi{10.1140/epjc/s10052-020-8341-8}.
\newblock \urlprefix\url{https://doi.org/10.1140/epjc/s10052-020-8341-8}

\bibitem{Spergel}
D.N. Spergel, P.J. Steinhardt, Observational evidence for self-interacting cold
  dark matter.
\newblock Phys. Rev. Lett. \textbf{84}, 3760--3763 (2000).
\newblock \doi{10.1103/PhysRevLett.84.3760}.
\newblock \urlprefix\url{https://link.aps.org/doi/10.1103/PhysRevLett.84.3760}

\bibitem{TULIN20181}
S.~Tulin, H.B. Yu, Dark matter self-interactions and small scale structure.
\newblock Physics Reports \textbf{730}, 1--57 (2018).
\newblock \doi{https://doi.org/10.1016/j.physrep.2017.11.004}.
\newblock
  \urlprefix\url{https://www.sciencedirect.com/science/article/pii/S0370157317304039}

\bibitem{Kaplinghat}
M.~Kaplinghat, S.~Tulin, H.B. Yu, Dark matter halos as particle colliders:
  Unified solution to small-scale structure puzzles from dwarfs to clusters.
\newblock Phys. Rev. Lett. \textbf{116}, 041,302 (2016).
\newblock \doi{10.1103/PhysRevLett.116.041302}.
\newblock
  \urlprefix\url{https://link.aps.org/doi/10.1103/PhysRevLett.116.041302}

\bibitem{Anand_2017}
S.~Anand, P.~Chaubal, A.~Mazumdar, S.~Mohanty, Cosmic viscosity as a remedy for
  tension between {PLANCK} and {LSS} data.
\newblock Journal of Cosmology and Astroparticle Physics \textbf{2017}(11),
  005--005 (2017).
\newblock \doi{10.1088/1475-7516/2017/11/005}.
\newblock \urlprefix\url{https://doi.org/10.1088/1475-7516/2017/11/005}

\bibitem{WiggleZDE}
M.I. Scrimgeour, T.~Davis, C.~Blake, J.B. James, G.B. Poole, L.~Staveley-Smith,
  S.~Brough, M.~Colless, C.~Contreras, W.~Couch, S.~Croom, D.~Croton, M.J.
  Drinkwater, K.~Forster, D.~Gilbank, M.~Gladders, K.~Glazebrook, B.~Jelliffe,
  R.J. Jurek, I.h. Li, B.~Madore, D.C. Martin, K.~Pimbblet, M.~Pracy, R.~Sharp,
  E.~Wisnioski, D.~Woods, T.K. Wyder, H.K.C. Yee, {The WiggleZ Dark Energy
  Survey: the transition to large-scale cosmic homogeneity}.
\newblock Monthly Notices of the Royal Astronomical Society \textbf{425}(1),
  116--134 (2012).
\newblock \doi{10.1111/j.1365-2966.2012.21402.x}.
\newblock \urlprefix\url{https://doi.org/10.1111/j.1365-2966.2012.21402.x}.
\newblock
  {\href{https://arxiv.org/abs/https://academic.oup.com/mnras/article-pdf/425/1/116/3176116/425-1-116.pdf}{{https://academic.oup.com/mnras/article-pdf/425/1/116/3176116/425-1-116.pdf}}}

\bibitem{Sylos_Labini_2009}
F.S. Labini, N.L. Vasilyev, L.~Pietronero, Y.V. Baryshev, Absence of
  self-averaging and of homogeneity in the large-scale galaxy distribution.
\newblock Europhysics Letters \textbf{86}(4), 49,001 (2009).
\newblock \doi{10.1209/0295-5075/86/49001}.
\newblock \urlprefix\url{https://dx.doi.org/10.1209/0295-5075/86/49001}

\bibitem{wiegand_scale}
A.~Wiegand, T.~Buchert, M.~Ostermann, {Direct Minkowski Functional analysis of
  large redshift surveys: a new high-speed code tested on the luminous red
  galaxy Sloan Digital Sky Survey-DR7 catalogue}.
\newblock Monthly Notices of the Royal Astronomical Society \textbf{443}(1),
  241--259 (2014).
\newblock \doi{10.1093/mnras/stu1118}.
\newblock \urlprefix\url{https://doi.org/10.1093/mnras/stu1118}.
\newblock
  {\href{https://arxiv.org/abs/https://academic.oup.com/mnras/article-pdf/443/1/241/4292640/stu1118.pdf}{{https://academic.oup.com/mnras/article-pdf/443/1/241/4292640/stu1118.pdf}}}

\bibitem{Shirokov1998}
M.F. Shirokov, I.Z. Fisher, Isotropic space with discrete gravitational-field
  sources. on the theory of a nonhomogeneous isotropic universe.
\newblock General Relativity and Gravitation \textbf{30}(9), 1411--1427 (1998).
\newblock \doi{10.1023/A:1018860826417}.
\newblock \urlprefix\url{https://doi.org/10.1023/A:1018860826417}

\bibitem{Ellis1984}
G.F.R. Ellis, \emph{Relativistic Cosmology: Its Nature, Aims and Problems}
  (Springer Netherlands, Dordrecht, 1984), pp. 215--288.
\newblock \doi{10.1007/978-94-009-6469-3_14}.
\newblock \urlprefix\url{https://doi.org/10.1007/978-94-009-6469-3_14}

\bibitem{Futamase}
T.~Futamase, Approximation scheme for constructing a clumpy universe in general
  relativity.
\newblock Phys. Rev. Lett. \textbf{61}, 2175--2178 (1988).
\newblock \doi{10.1103/PhysRevLett.61.2175}.
\newblock \urlprefix\url{https://link.aps.org/doi/10.1103/PhysRevLett.61.2175}

\bibitem{Zalaletdinov1992}
R.M. Zalaletdinov, Averaging out the einstein equations.
\newblock General Relativity and Gravitation \textbf{24}(10), 1015--1031
  (1992).
\newblock \doi{10.1007/BF00756944}.
\newblock \urlprefix\url{https://doi.org/10.1007/BF00756944}

\bibitem{Zalaletdinov1993}
R.M. Zalaletdinov, Towards a theory of macroscopic gravity.
\newblock General Relativity and Gravitation \textbf{25}(7), 673--695 (1993).
\newblock \doi{10.1007/BF00756937}.
\newblock \urlprefix\url{https://doi.org/10.1007/BF00756937}

\bibitem{Buchert}
T.~Buchert, On average properties of inhomogeneous fluids in general
  relativity: Dust cosmologies.
\newblock General Relativity and Gravitation \textbf{32}(1), 105--125 (2000).
\newblock \doi{10.1023/A:1001800617177}.
\newblock \urlprefix\url{https://doi.org/10.1023/A:1001800617177}

\bibitem{Weigand_et_al}
A.~Wiegand, T.~Buchert, Multiscale cosmology and structure-emerging dark
  energy: A plausibility analysis.
\newblock Phys. Rev. D \textbf{82}, 023,523 (2010).
\newblock \doi{10.1103/PhysRevD.82.023523}.
\newblock \urlprefix\url{https://link.aps.org/doi/10.1103/PhysRevD.82.023523}

\bibitem{Schwarz}
D.J. Schwarz.
\newblock Accelerated expansion without dark energy (2002).
\newblock \doi{10.48550/ARXIV.ASTRO-PH/0209584}.
\newblock \urlprefix\url{https://arxiv.org/abs/astro-ph/0209584}

\bibitem{Rasanen_2004}
S.~Räsänen, Dark energy from back-reaction.
\newblock Journal of Cosmology and Astroparticle Physics \textbf{2004}(02),
  003--003 (2004).
\newblock \doi{10.1088/1475-7516/2004/02/003}.
\newblock \urlprefix\url{https://doi.org/10.1088/1475-7516/2004/02/003}

\bibitem{wiltshire}
D.L. Wiltshire, \emph{Dark Energy without Dark Energy} (World Scientific,
  2008), pp. 565--596.
\newblock \doi{10.1142/9789812814357_0053}.
\newblock
  \urlprefix\url{https://www.worldscientific.com/doi/abs/10.1142/$9789812814357_0053$}

\bibitem{Kolb_2006}
E.W. Kolb, S.~Matarrese, A.~Riotto, On cosmic acceleration without dark energy.
\newblock New Journal of Physics \textbf{8}(12), 322--322 (2006).
\newblock \doi{10.1088/1367-2630/8/12/322}.
\newblock \urlprefix\url{https://doi.org/10.1088/1367-2630/8/12/322}

\bibitem{Ishibashi_2005}
A.~Ishibashi, R.M. Wald, Can the acceleration of our universe be explained by
  the effects of inhomogeneities?
\newblock Classical and Quantum Gravity \textbf{23}(1), 235--250 (2005).
\newblock \doi{10.1088/0264-9381/23/1/012}.
\newblock \urlprefix\url{https://doi.org/10.1088/0264-9381/23/1/012}

\bibitem{Riess_2021}
A.G. Riess, S.~Casertano, W.~Yuan, J.B. Bowers, L.~Macri, J.C. Zinn,
  D.~Scolnic, Cosmic distances calibrated to 1$\%$ precision with gaia edr3
  parallaxes and hubble space telescope photometry of 75 milky way cepheids
  confirm tension with $\lambda$cdm.
\newblock The Astrophysical Journal Letters \textbf{908}(1), L6 (2021).
\newblock \doi{10.3847/2041-8213/abdbaf}.
\newblock \urlprefix\url{https://dx.doi.org/10.3847/2041-8213/abdbaf}

\bibitem{Freedman_2021}
W.L. Freedman, Measurements of the hubble constant: Tensions in perspective*.
\newblock The Astrophysical Journal \textbf{919}(1), 16 (2021).
\newblock \doi{10.3847/1538-4357/ac0e95}.
\newblock \urlprefix\url{https://dx.doi.org/10.3847/1538-4357/ac0e95}

\bibitem{Heinesen_2020}
A.~Heinesen, T.~Buchert, Solving the curvature and hubble parameter
  inconsistencies through structure formation-induced curvature.
\newblock Classical and Quantum Gravity \textbf{37}(16), 164,001 (2020).
\newblock \doi{10.1088/1361-6382/ab954b}.
\newblock \urlprefix\url{https://dx.doi.org/10.1088/1361-6382/ab954b}

\bibitem{Coley}
A.A. Coley.
\newblock Null geodesics and observational cosmology (2008).
\newblock \doi{10.48550/ARXIV.0812.4565}.
\newblock \urlprefix\url{https://arxiv.org/abs/0812.4565}

\bibitem{R_s_nen_2009}
S.~Räsänen, Light propagation in statistically homogeneous and isotropic dust
  universes.
\newblock Journal of Cosmology and Astroparticle Physics \textbf{2009}(02),
  011--011 (2009).
\newblock \doi{10.1088/1475-7516/2009/02/011}.
\newblock \urlprefix\url{https://doi.org/10.1088/1475-7516/2009/02/011}

\bibitem{Gasperini_2011}
M.~Gasperini, G.~Marozzi, F.~Nugier, G.~Veneziano, Light-cone averaging in
  cosmology: formalism and applications.
\newblock Journal of Cosmology and Astroparticle Physics \textbf{2011}(07),
  008--008 (2011).
\newblock \doi{10.1088/1475-7516/2011/07/008}.
\newblock \urlprefix\url{https://doi.org/10.1088/1475-7516/2011/07/008}

\bibitem{Fleury}
P.~Fleury, H.~Dupuy, J.P. Uzan, Interpretation of the hubble diagram in a
  nonhomogeneous universe.
\newblock Phys. Rev. D \textbf{87}, 123,526 (2013).
\newblock \doi{10.1103/PhysRevD.87.123526}.
\newblock \urlprefix\url{https://link.aps.org/doi/10.1103/PhysRevD.87.123526}

\bibitem{Fleury2}
P.~Fleury, H.~Dupuy, J.P. Uzan, Can all cosmological observations be accurately
  interpreted with a unique geometry?
\newblock Phys. Rev. Lett. \textbf{111}, 091,302 (2013).
\newblock \doi{10.1103/PhysRevLett.111.091302}.
\newblock
  \urlprefix\url{https://link.aps.org/doi/10.1103/PhysRevLett.111.091302}

\bibitem{Fleury3_2014}
P.~Fleury, Swiss-cheese models and the dyer-roeder approximation.
\newblock Journal of Cosmology and Astroparticle Physics \textbf{2014}(06),
  054--054 (2014).
\newblock \doi{10.1088/1475-7516/2014/06/054}.
\newblock \urlprefix\url{https://doi.org/10.1088/1475-7516/2014/06/054}

\bibitem{Bagheri_2014}
S.~Bagheri, D.J. Schwarz, Light propagation in the averaged universe.
\newblock Journal of Cosmology and Astroparticle Physics \textbf{2014}(10),
  073--073 (2014).
\newblock \doi{10.1088/1475-7516/2014/10/073}.
\newblock \urlprefix\url{https://doi.org/10.1088/1475-7516/2014/10/073}

\bibitem{Futamase2}
T.~Futamase, M.~Sasaki, Light propagation and the distance-redshift relation in
  a realistic inhomogeneous universe.
\newblock Phys. Rev. D \textbf{40}, 2502--2510 (1989).
\newblock \doi{10.1103/PhysRevD.40.2502}.
\newblock \urlprefix\url{https://link.aps.org/doi/10.1103/PhysRevD.40.2502}

\bibitem{Koksbang_2019}
S.~Koksbang, Another look at redshift drift and the backreaction conjecture.
\newblock Journal of Cosmology and Astroparticle Physics \textbf{2019}(10),
  036–036 (2019).
\newblock \doi{10.1088/1475-7516/2019/10/036}.
\newblock \urlprefix\url{http://dx.doi.org/10.1088/1475-7516/2019/10/036}

\bibitem{Koksbang2}
S.~Koksbang, {Observations in statistically homogeneous, locally inhomogeneous
  cosmological toy models without FLRW backgrounds}.
\newblock Monthly Notices of the Royal Astronomical Society: Letters
  \textbf{498}(1), L135--L139 (2020).
\newblock \doi{10.1093/mnrasl/slaa146}.
\newblock \urlprefix\url{https://doi.org/10.1093/mnrasl/slaa146}.
\newblock
  {\href{https://arxiv.org/abs/https://academic.oup.com/mnrasl/article-pdf/498/1/L135/33718579/slaa146.pdf}{{https://academic.oup.com/mnrasl/article-pdf/498/1/L135/33718579/slaa146.pdf}}}

\bibitem{Koksbang_PRL}
S.M. Koksbang, Searching for signals of inhomogeneity using multiple probes of
  the cosmic expansion rate $h(z)$.
\newblock Phys. Rev. Lett. \textbf{126}, 231,101 (2021).
\newblock \doi{10.1103/PhysRevLett.126.231101}.
\newblock
  \urlprefix\url{https://link.aps.org/doi/10.1103/PhysRevLett.126.231101}

\bibitem{R_s_nen_2008}
S.~Räsänen, Evaluating backreaction with the peak model of structure
  formation.
\newblock Journal of Cosmology and Astroparticle Physics \textbf{2008}(04), 026
  (2008).
\newblock \doi{10.1088/1475-7516/2008/04/026}.
\newblock \urlprefix\url{https://doi.org/10.1088/1475-7516/2008/04/026}

\bibitem{bose}
N.~Bose, A.S. Majumdar, {Future deceleration due to cosmic backreaction in
  presence of the event horizon}.
\newblock Monthly Notices of the Royal Astronomical Society: Letters
  \textbf{418}(1), L45--L48 (2011).
\newblock \doi{10.1111/j.1745-3933.2011.01140.x}.
\newblock \urlprefix\url{https://doi.org/10.1111/j.1745-3933.2011.01140.x}.
\newblock
  {\href{https://arxiv.org/abs/https://academic.oup.com/mnrasl/article-pdf/418/1/L45/6416897/418-1-L45.pdf}{{https://academic.oup.com/mnrasl/article-pdf/418/1/L45/6416897/418-1-L45.pdf}}}

\bibitem{Bose2013}
N.~Bose, A.S. Majumdar, Effect of cosmic backreaction on the future evolution
  of an accelerating universe.
\newblock General Relativity and Gravitation \textbf{45}(10), 1971--1987
  (2013).
\newblock \doi{10.1007/s10714-013-1572-3}.
\newblock \urlprefix\url{https://doi.org/10.1007/s10714-013-1572-3}

\bibitem{Ali_2017}
A.~Ali, A.~Majumdar, Future evolution in a backreaction model and the analogous
  scalar field cosmology.
\newblock Journal of Cosmology and Astroparticle Physics \textbf{2017}(01),
  054–054 (2017).
\newblock \doi{10.1088/1475-7516/2017/01/054}.
\newblock \urlprefix\url{http://dx.doi.org/10.1088/1475-7516/2017/01/054}

\bibitem{Pandey_2022}
S.S. Pandey, A.~Sarkar, A.~Ali, A.~Majumdar, Effect of inhomogeneities on the
  propagation of gravitational waves from binaries of compact objects.
\newblock Journal of Cosmology and Astroparticle Physics \textbf{2022}(06), 021
  (2022).
\newblock \doi{10.1088/1475-7516/2022/06/021}.
\newblock \urlprefix\url{https://doi.org/10.1088/1475-7516/2022/06/021}

\bibitem{Weinberg}
S.~Weinberg, \emph{Gravitation and Cosmology: Principles and Applications of
  the General Theory of Relativity} (John Wiley $\&$ Sons, Inc., 1972)

\bibitem{Pimentel2016}
O.M. Pimentel, F.D. Lora-Clavijo, G.A. Gonz{\'a}lez, The energy-momentum tensor
  for a dissipative fluid in general relativity.
\newblock General Relativity and Gravitation \textbf{48}(10), 124 (2016).
\newblock \doi{10.1007/s10714-016-2121-7}.
\newblock \urlprefix\url{https://doi.org/10.1007/s10714-016-2121-7}

\bibitem{Barbosa_2017}
C.M.S. Barbosa, H.~Velten, J.C. Fabris, R.O. Ramos, Assessing the impact of
  bulk and shear viscosities on large scale structure formation.
\newblock Phys. Rev. D \textbf{96}, 023,527 (2017).
\newblock \doi{10.1103/PhysRevD.96.023527}.
\newblock \urlprefix\url{https://link.aps.org/doi/10.1103/PhysRevD.96.023527}

\bibitem{Velten_2014}
H.~Velten, T.R.P. Caram\^es, J.C. Fabris, L.~Casarini, R.C. Batista, Structure
  formation in a $\mathrm{\ensuremath{\Lambda}}$ viscous cdm universe.
\newblock Phys. Rev. D \textbf{90}, 123,526 (2014).
\newblock \doi{10.1103/PhysRevD.90.123526}.
\newblock \urlprefix\url{https://link.aps.org/doi/10.1103/PhysRevD.90.123526}

\bibitem{R_s_nen_2006}
S.~Räsänen, Accelerated expansion from structure formation.
\newblock Journal of Cosmology and Astroparticle Physics \textbf{2006}(11),
  003--003 (2006).
\newblock \doi{10.1088/1475-7516/2006/11/003}.
\newblock \urlprefix\url{https://doi.org/10.1088/1475-7516/2006/11/003}

\bibitem{brevik3}
B.D. Normann, I.~Brevik, Characteristic properties of two different viscous
  cosmology models for the future universe.
\newblock Modern Physics Letters A \textbf{32}(04), 1750,026 (2017).
\newblock \doi{10.1142/S0217732317500262}.
\newblock \urlprefix\url{https://doi.org/10.1142/S0217732317500262}.
\newblock
  {\href{https://arxiv.org/abs/https://doi.org/10.1142/S0217732317500262}{{https://doi.org/10.1142/S0217732317500262}}}

\bibitem{brevik_entropy}
B.D. Normann, I.~Brevik, General bulk-viscous solutions and estimates of bulk
  viscosity in the cosmic fluid.
\newblock Entropy \textbf{18}(6) (2016).
\newblock \doi{10.3390/e18060215}.
\newblock \urlprefix\url{https://www.mdpi.com/1099-4300/18/6/215}

\bibitem{wang_meng}
J.~WANG, X.~MENG, Effects of new viscosity model on cosmological evolution.
\newblock Modern Physics Letters A \textbf{29}(03), 1450,009 (2014).
\newblock \doi{10.1142/S0217732314500096}.
\newblock \urlprefix\url{https://doi.org/10.1142/S0217732314500096}.
\newblock
  {\href{https://arxiv.org/abs/https://doi.org/10.1142/S0217732314500096}{{https://doi.org/10.1142/S0217732314500096}}}

\bibitem{CUSAT_Sasidharan2016}
A.~Sasidharan, T.K. Mathew, Phase space analysis of bulk viscous matter
  dominated universe.
\newblock Journal of High Energy Physics \textbf{2016}(6), 138 (2016).
\newblock \doi{10.1007/JHEP06(2016)138}.
\newblock \urlprefix\url{https://doi.org/10.1007/JHEP06(2016)138}

\bibitem{Velten_PRD}
H.~Velten, D.J. Schwarz, Dissipation of dark matter.
\newblock Phys. Rev. D \textbf{86}, 083,501 (2012).
\newblock \doi{10.1103/PhysRevD.86.083501}.
\newblock \urlprefix\url{https://link.aps.org/doi/10.1103/PhysRevD.86.083501}

\bibitem{R_s_nen_4_2010}
S.~Räsänen, Light propagation in statistically homogeneous and isotropic
  universes with general matter content.
\newblock Journal of Cosmology and Astroparticle Physics \textbf{2010}(03),
  018--018 (2010).
\newblock \doi{10.1088/1475-7516/2010/03/018}.
\newblock \urlprefix\url{https://doi.org/10.1088/1475-7516/2010/03/018}

\bibitem{maggiore}
M.~Maggiore, \emph{{Gravitational Waves: Volume 1: Theory and Experiments}}
  (Oxford University Press, 2007).
\newblock \doi{10.1093/acprof:oso/9780198570745.001.0001}.
\newblock
  \urlprefix\url{https://doi.org/10.1093/acprof:oso/9780198570745.001.0001}

\bibitem{viscosity_lepton}
L.~Husdal, Viscosity in a lepton-photon universe.
\newblock Astrophysics and Space Science \textbf{361}(8), 263 (2016).
\newblock \doi{10.1007/s10509-016-2847-4}.
\newblock \urlprefix\url{https://doi.org/10.1007/s10509-016-2847-4}

\bibitem{Rosado}
P.A. Rosado, P.D. Lasky, E.~Thrane, X.~Zhu, I.~Mandel, A.~Sesana, Detectability
  of gravitational waves from high-redshift binaries.
\newblock Phys. Rev. Lett. \textbf{116}, 101,102 (2016).
\newblock \doi{10.1103/PhysRevLett.116.101102}.
\newblock
  \urlprefix\url{https://link.aps.org/doi/10.1103/PhysRevLett.116.101102}

\end{thebibliography}
\end{document}